\documentclass[a4paper]{article}

\usepackage{amssymb} 
\usepackage{amstext} 
\usepackage{amsmath}

\usepackage[english]{babel}
\usepackage[T1]{fontenc}
\usepackage{lmodern}
\usepackage[latin1]{inputenc}
\usepackage{graphicx}
\usepackage{natbib}
\usepackage{breakcites}

\newcommand{\W}[2]{
W^{\scriptscriptstyle{#2}}_{\scriptscriptstyle{#1}}
}

\newcounter{figno}
\newcommand{\figlabel}[1]{\refstepcounter{figno}\label{#1}}

\def\Authors{Steve N'Guyen\,$^{1,2*}$, Charles Thurat\,$^{1}$ and
  Beno{\^i}t Girard\,$^1$}
\def\Address{$^{1}$ Institut des Syst{\`e}mes Intelligents et de
  Robotique\\ Universit{\'e} Pierre et Marie Curie-Paris 6\\ CNRS UMR 7222, Paris, France \\
$^{2}$ LPPA, Coll{\`e}ge de France, CNRS UMR 7152, Paris, France}

\date{}
\begin{document}

\title{Saccade learning with concurrent cortical and subcortical basal ganglia loops}
\author{\Authors\\ \Address}

\maketitle
\begin{abstract}

\section*{}

The Basal Ganglia is a central structure involved in multiple cortical
and subcortical loops. Some of these loops are believed to be
responsible for saccade target selection. We study here how the very
specific structural relationships of these saccadic loops can
affect the ability of learning spatial and feature-based tasks.

We propose a model of saccade generation with
reinforcement learning capabilities based on
our previous basal ganglia and superior colliculus models.
It is structured around the interactions of two parallel cortico-basal
loops and one tecto-basal loop.
The two cortical loops
separately deal with spatial and non-spatial information to
select targets in a concurrent way.  The subcortical loop is used to
make the final target selection leading to the production of the
saccade.  These different loops may work in concert or disturb each
other regarding reward maximization.  Interactions between these loops
and their learning capabilities are tested on different saccade tasks.

The results show the ability of this model to correctly learn basic target
selection based on different criteria (spatial or not).
Moreover the model reproduces and explains training dependent express
saccades toward targets based on a spatial criterion.    

Finally, the model predicts that in absence of prefrontal control, the
spatial loop should dominate.

\paragraph{Keywords:}
basal ganglia, superior colliculus, saccades, decision making, reinforcement learning 
\end{abstract}

\section{Introduction}

The basal ganglia (BG) are a set of interconnected subcortical nuclei
\citep{redgrave07}, which are thought to be central in the performance
of action selection \citep{mink96,redgrave99}.

\begin{figure}
\centering
  \includegraphics[width=\linewidth]{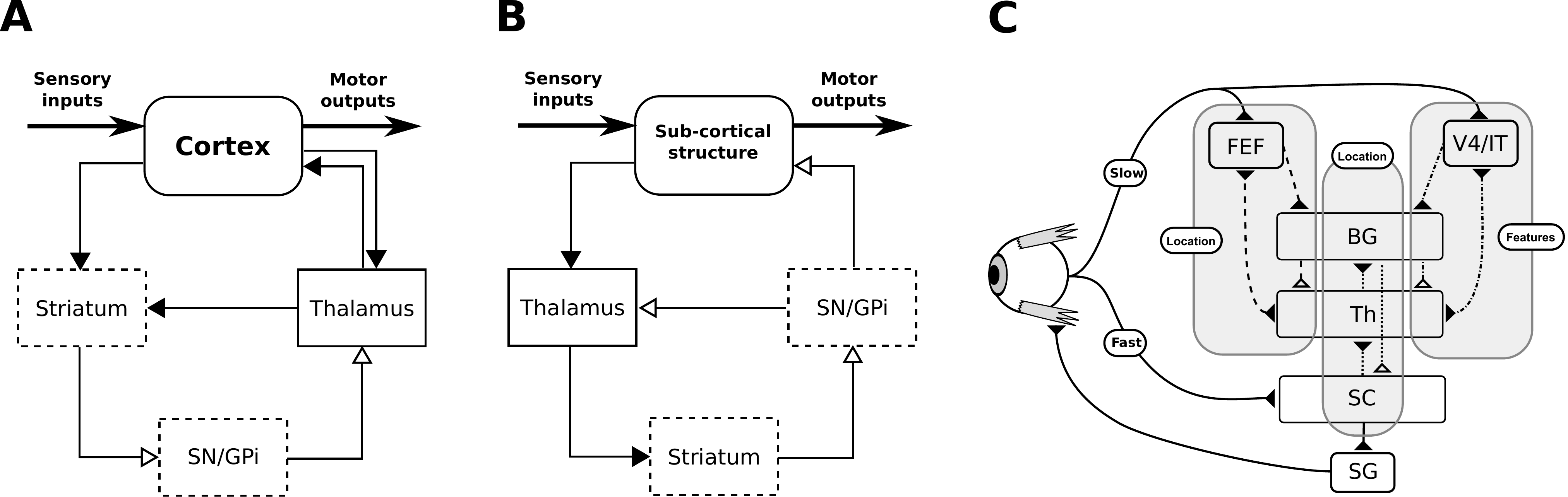}
  \caption{\label{bgloops}
    A: general organization for cortical loops. B: general
    organization for subcortical loops. Filled arrow heads are
    exitatory connexions, empty arrow heads are inhibitory
    connexions. Dashed block are inhibitory structures. Note that the
    concerned thalamus nuclei are differents between A (ventral
    anterior, ventrolateral, medial dorsal) and B (pulvinar, lateral
    posterior, rostral and caudal intralaminar). A and B adapted from
    \citep{McHaffie2005}. C: schematic representation of the
    relationships between the three modelled loops, note the type of
    information processed (either location or features of targets) and
    the delays (slow or fast).
}
  
\end{figure}

The BG are traditionally described as being composed of various
parallel subcircuits with identical internal wiring, implied in
different functions (from motor to cognitive ones), and belonging to a
set of parallel cortico-baso-thalamo-cortical loops
\citep{alexander86}, as schematized in Fig.~\ref{bgloops}A. However,
the BG also participate in purely subcortical loops
\citep{Groenewegen1994,McHaffie2005,McHaffie&al2006,May2006}, which
are wired a bit differently as the input to the BG is relayed through
the thalamus and the BG output projects directly to the considered
subcortical structures (Fig.~\ref{bgloops}B), and which rely on
different thalamic nuclei (pulvinar, lateral posterior, rostral and
caudal intralaminar). They do, in particular, participate in loops
with the superior colliculus (SC), well-known for its laminar
structure, its mapping of the visual field and its involvement in gaze
orientation movements, including saccadic eye movements
\citep{Moschovakis1996,LynchTian2006}.


We propose here a computational model of the interactions of
subcortical and cortical BG loops in primates, processing either
target position (spatial) information or target feature information,
in the well investigated framework of saccadic eye movements
\citep{Hikosaka2000a}. Indeed, cortico-basal loops dealing with the
location of potential targets in the visual field, on the one hand, or
with the detection of features of potential targets, on the other
hand, have long been identified. The superior colliculus (and thus the
tecto-basal loop) is a bottleneck receiving all this information
for the final decision, however it also receives target location
information earlier than the cortically processed information, through
direct projections from the retina.  

We thus study the effects imposed by this hierarchical structure
-- where the highest level modules have longer latencies, while the
lowest level module has a lower latency shortcut, but specific to
location information, Fig.~\ref{bgloops}C -- on performance and saccadic
reaction time in space-based and/or feature-based selection tasks, in
order to identify predictions specific to this organization. These
predictions stand for dorsolateral prefrontal cortex (dlPFC) deprived
animals as it is not included in our model and as we can expect the
inhibitory control from the dlPFC on the superior colliculus to allow
additional control on unwanted short-latency saccades \citep{koval11}.

We show that the fact that a purely spatial selection and learning
system operate at the last level predicts that:
\begin{itemize}
  \item in spatial tasks only should the saccadic reaction times
    decrease with learning, allowing the generation of express
    saccades and causing short latency activations in the FEF,
  \item performance in feature-based tasks should be lower than in
    spatial tasks, because of the perturbations caused by the
    subcortical spatial loop,
  \item in conjunction tasks, where spatial and feature-based
    information determine the good choice, errors are unavoidable when
    no choice should be made. 

\end{itemize}

\section{Material \& Methods}\label{sec:Methods}

\subsection{Global architecture}

The subcortical loop (Fig.~\ref{model}, dotted circuit) has access to
visual inputs directly conveyed from the retina to the superficial
layers of the superior colliculus, with a low
latency. These retinal projections provide relatively rich visual information
\citep{Girman2007}, but no color information. As the SC layers are
organized as piled retinotopic maps of the visual field, and given the
spatial receptive fields of the BG output neurons projecting to the SC
\citep{hikosaka1983visualIV}, it can be assumed that the competition
among targets is here based on spatial position.  This loop is a good
candidate neural substrate to explain the accumulating evidence (see
for example \citet{McPeek2002,McPeek2003,McPeek2004}, among many
others since 2000) that the SC performs target selection on its own,
rather than solely executing cortical decisions.  

Two cortical loops, projecting to the SC as a common output, are
considered. A first one (Fig.~\ref{model}, dashed circuit), comprising
the frontal eye fields (FEF), also operates on the spatial domain, but
contributes to saccade generation with longer latencies than the
SC. This loop is known to be a common pathway for ``cognitive''
saccades, where working memory or sequence generation are involved,
however these are not included in the proposed model (indeed no SEF
and pre-SEF have been included). We hypothesize that the BG subcircuit
involved in this loop is shared with the subcortical one (i.e. there
is only one BG subcircuit dedicated to spatial selection of targets).
This choice of converging input has been made based on known anatomy
as it seems that FEF projects to the ``Oculomotor
Striatum'' (central/longitudinal Caudate) \citep{Stanton1988}.

The second one (Fig.~\ref{model}, dash-dot circuit) comprises V4 and
IT and deals with the selection of targets exhibiting specific
features (only color will be used here for simplicity).  V4 is known
to be selective to shape and color \citep{Ogawa2004} and
visuotopically organized \citep{Gattass1988}. Moreover, this region
exhibits strong recurrent connections with IT (in particular the TE
area) \citep{Ungerleider2008}. The TE region of IT has been shown to
be selective to features (and colors) and not visuotopically organized
(activity doesn't depend on object position) \citep{Tompa2010}. More
importantly, this TE area forms a loop with the Basal Ganglia
\citep{Middleton1996a}, thus it seems somewhat reasonable to
hypothesize that colors and features could be selected through a
cortical IT-BG-Th loop in a non-spatial fashion and then projected
back to V4.
In particular, the TE region, projects to the ``Visual Striatum''
(tail of Caudate and caudal/ventral portion of Putamen)
\citep{Middleton1996a}, supporting the separation between the spatial
and the feature loop.
The Superior Colliculus is known to receive numerous projections from
cortical areas amongst which V4 \citep{Fries1984,Lock2003}.  This
mechanism is compatible with feature/color sensitivity with a longer
latency than luminance signal observed in intermediate layers of SC
(SCi) \citep{White2009,White2011}.

So to summarize, in this model two parallel mechanisms compete for target
selection (Fig.~\ref{bgloops}, C).
The first one is ``location'' based and comprises two cooperating
loops, both cortical and subcortical.
The second one is ``feature'' based and comprises one cortical loop.
The detail of the equations are given in section \ref{sec:Methods}.

\subsection{Model description}

\begin{figure}
\centering
  \includegraphics[width=0.7\linewidth]{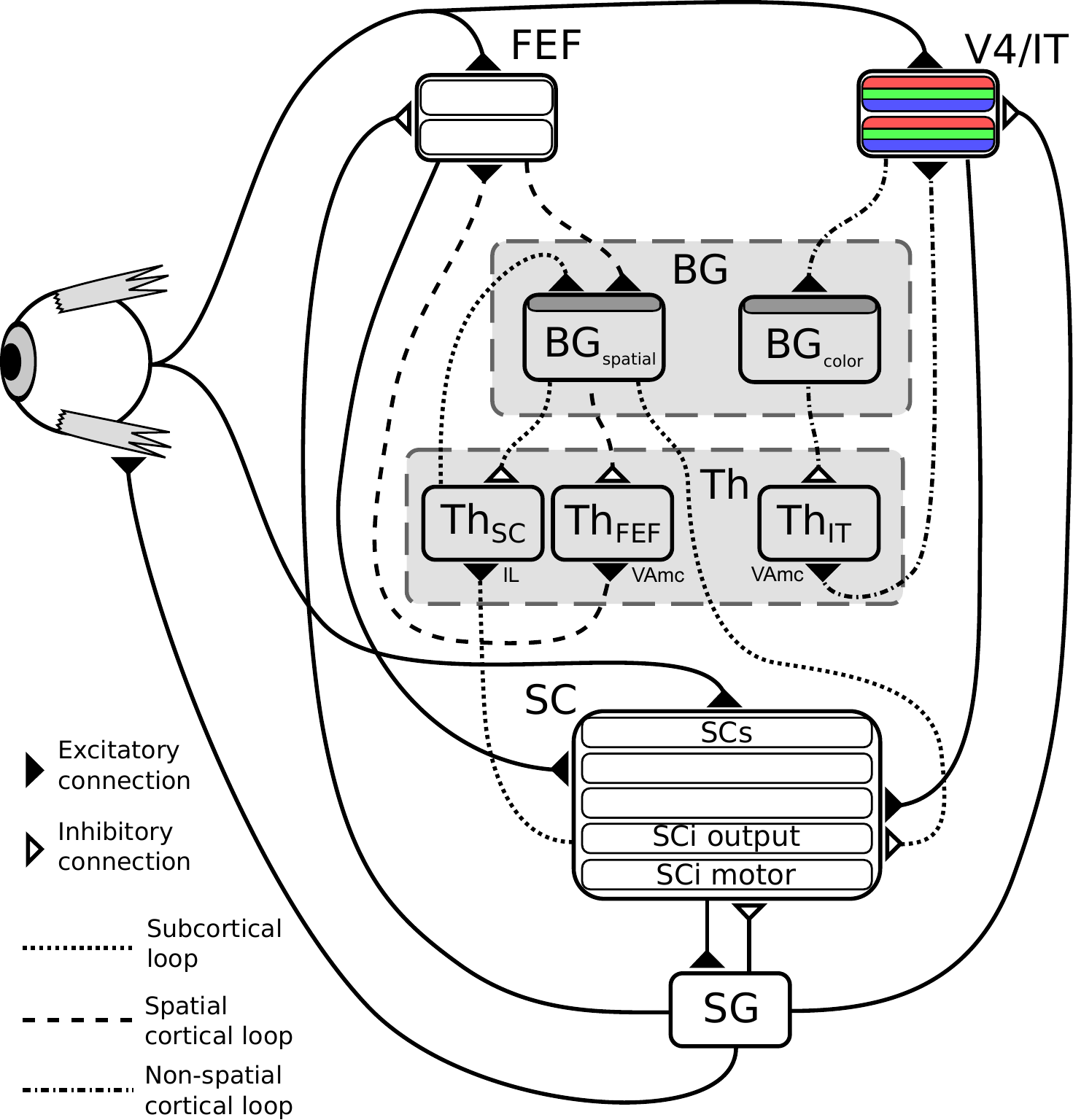}
  \caption{\label{model}
Structure of the model. BG: Basal ganglia;
    FEF: frontal eye fields; SG: saccade generators; SC: superior
    colliculus; Th: thalamus; V4\textbar IT: Feature perception area including
    IT (TE region) interacting with V4 visual cortex area. Dark gray
    shaded layers on BG modules are input layers with reinforcement learning
    capabilities.}
    
\end{figure}

The proposed model is intended to learn to generate saccades towards
targets selected based on their color and location in the visual field
(cf. Fig.~\ref{model}), depending on the reward contingencies
experienced during interaction with the environment.


As said before, it is composed of three main loops going through the
basal ganglia, which interact in both competitive and cooperative
ways. The subcortical one corresponds to the SC-Th-BG circuit (dotted
connexions on Fig.~\ref{model}), it gets its inputs from the direct
projections from the retina to the superficial layers of the superior
colliculus along with activity of deep layers, and it selects among
targets competing on a purely spatial dimension.  This loop passes
through the Intralaminar nucleus (IL) thalamic relay
\citep{McHaffie2005}.

The cortical ones also comprise a circuit dedicated to spatial
competition (FEF-BG-Th, dashed connexions on Fig.~\ref{model}), which
shares its BG circuit with the subcortical loop but with a different
thalamic relay (the paralamellar portion of the
mediodorsal thalamic nuclei, MDpl) \citep{Alexander1986,Tian1997}
and another dedicated
to features (namely color) selection (IT-BG-Th, dot-dashed connexions
on Fig.~\ref{model}) via VAmc \citep{Middleton1996a}.

Retinal information is transmitted to SC, FEF and V4\textbar IT with different
latencies according to the literature,
SCs input latency is fixed to 41~ms (type I neurons) \citep{Rizzolatti1980}. 
FEF to 91~ms and IT to 122~ms (average over all TE sub-regions) \citep{Lamme2000}.

The FEF module contains an input and an output retinotopic map sensible to luminance.
The V4\textbar IT module contains one input and one output retinotopic maps for
each color.  
SC module also contains several retinotopic maps, dealing
with direct retinal input (SCs), FEF input, V4\textbar IT input,
summed activity of SCs, FEF and V4\textbar IT (SCi output) and
motor activity (SCi motor).
For each of these structures a selection loop through BG occurs.

We use rate-coding models of neurons (based on locally projected
dynamical systems, lPDS, \citep{Girard2008}), which are defined as
follows:

\begin{equation}
 \dot{x} = \Pi_{[0,max]}(x(t), \frac{I(t)- x(t)}{\tau})
 \label{eq:NNEquation}
\end{equation}

where $I(t)$ represents the external inputs, $\tau$ the time constant,
and $\Pi_{[0,max]}$ a projection operator ensuring that the neuron
activity $x(t)$ will remain within $[0,max]$.

The projection operator $\Pi_{[0,max]}$ is simply an operator acting
on $\dot{x}$ ensuring that the variable $x$ remains within a specified
range of values.
In our case (Euler integration with 1ms timestep) we end up with a
discrete update operating as follows :
\begin{equation}
x(t+dt) = min\big [ 1, max \big [ 0, x(t) + \frac{dt}{\tau} \times
(I(t)-x(t))\big ] \big ]
\end{equation}
This method is very similar to the classical way of converting the
computed activity $x$ into a non-negative one $y=max(0.0,x)$ but here the
non linear ``transfer function'' is applied inside the
differential equation at the cost of making it a non longer a
classical ordinary differential equation but with some over benefits
such as ``contraction'' i.e. stability.

The basal ganglia model
we use here \citep{Girard2008} was formulated in this framework, so as
to formally ensure its dynamical stability. For the sake of
consistency, we thus use it for the rest of the model presented here.
Only the external input part ($I(t)$) and the time constant ($\tau$)
of this equation have to be specified to define such a neuron
model. Thus, to simplify the writing, only $I(t)$ will be given in the
next section providing a detailed description of the model, while the
time constants and other model parameters are provided in supplemental
data section.

The BG exert an inhibitory influence on their target circuits, which
prevents them from generating actions. Even without any inputs, the BG
converge to a given level of inhibition, $GPi|SNr_{rest}$, sufficient
to enforce this control. As previously proposed in
\citep{Arai94,das1996,Arai1999}, we modeled the effect of the basal
ganglia inhibition as modulating the excitatory inputs of the targeted
systems. To ensure that, at rest, no action can be generated, this
inhibitory gain modulation is normalized with regards to the $GPi|SNr_{rest}$
constant. Thus, the contributions of the BG outputs to the circuits
they target will take the general following form in the equations of
the next section:
\begin{equation}
  W_{E} \times I_E \times (1 - \frac{GPi|SNr}{GPi|SNr_{rest}})
\end{equation}
Where $I_E$ is the excitatory input controlled by the BG inhibition,
$GPi|SNr$ is the output of the BG neurons projecting to the considered
circuit.

The feedback from the superior colliculus, which signals the
end of the execution of a saccade, is also modeled as modulating.

Most of the components of the model are $70 \times 70$ 2D maps of lPDS
neurons for each hemifield, respecting the complex-logarithmic
geometry of the macaque superior colliculus, as modeled by
\citep{Ottes1986}. Unless specified, neurons of one map project to
those of another map in a one-to-one manner.
Visual inputs are simulated as gaussian activities spreading over a
hundred of neurons.

\subsubsection{Cortical and subcortical loops\label{sec:loops}}

\begin{figure}
\centering
  \includegraphics[width=0.7\linewidth]{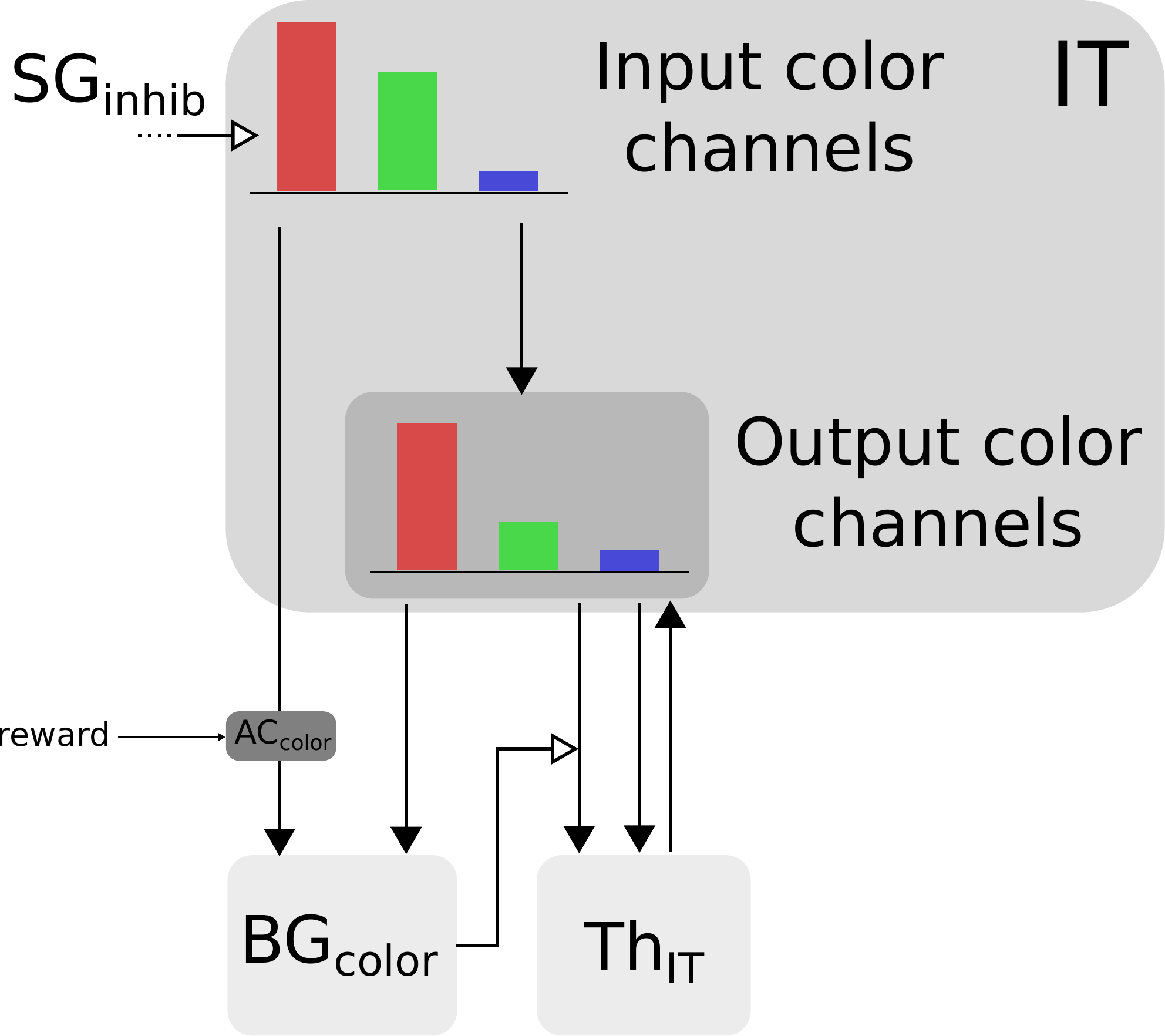}
  \caption{\label{V4} Selection loop for color channels. Black arrow heads
    are exitatory connexions, empty arrow heads are inhibitory connexions.}
\end{figure}

Color information is processed by the cortical V4\textbar IT-BG-Th loop. As
stated previously, the V4 structure contains several
retinotopic maps each encoding for a specific color (3 are used here,
red, green and blue).  In order to deal with non spatial color
channels, activity in each map is summed, providing a reduced number
of independent channel.  These channels are amplified in an closed
loop manner by the interaction of IT with BG and Th
(cf. Fig.~\ref{V4}).

\begin{figure}
\centering
  \includegraphics[width=1\linewidth]{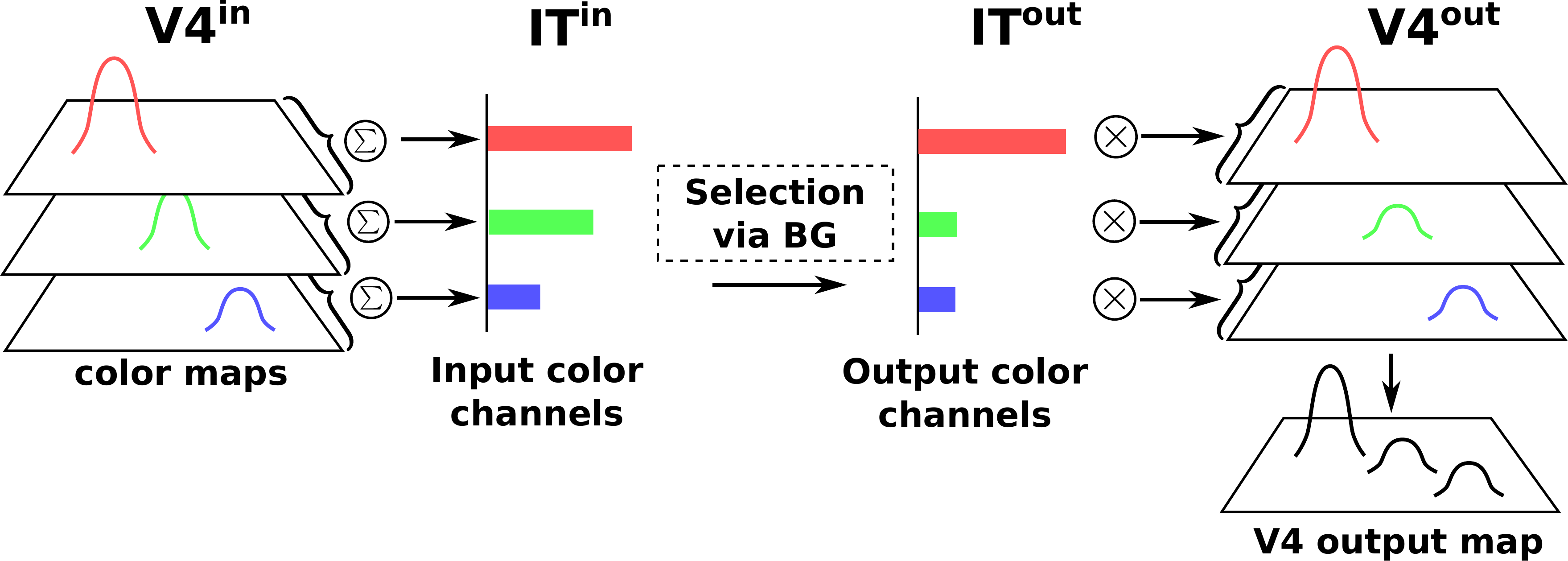}
  \caption{\label{featloc} Spatial-color transformation.}
\end{figure}

Thus, the BG selection occurring in the V4\textbar IT-BG-Th loop deals with
non-spatial color information only.  Then these channels are transformed
back into retinotopic maps (cf. Fig.~\ref{featloc}) and the resulting map
(V4 output map) is then projected to SCi.  Activity fed to the
channels is computed as follows:

\begin{align}\label{eq:V4_loop}
  IT_c^{out} =&
  \W{IT^{in}}{IT^{out}}.IT_c^{in}\times
  (1-W_{SG_{inhib}}.SG_{inhib})\notag\\
  &+\W{Th^{IT}}{IT^{out}}.Th_c^{IT} \\
  Th_c^{IT} =& IT_c^{out}\times \Big(
    \W{IT^{out}}{Th^{IT}}+\W{GPi}{Th^{IT}} \times (1-\frac{GPi|SNr_c^{color}}{GPi|SNr_{rest}})
    \Big) \notag\\
    &-\W{TRN^{IT}}{Th^{IT}}.TRN^{IT}+I_{Th}
\end{align}

with $c \in [red, green, blue]$, $IT_c^{in}$ the visual input channel
for color $c$, $IT_c^{out}$ the activity of the $IT$ layer connected
with $Th^{IT}$ and $SG_{inhib}$ the ascending inhibition from saccade
generators.  Thalamic activity depends on $IT_c^{out}$ and on BG
output nuclei $GPi|SNr_{color}$. The BG output thus gates a part of
the transmission between $IT^{out}$ and $Th^{IT}$ with a modulating
inhibition. $TRN^{IT}$ is the activity of the globally inhibiting
inputs from the thalamic reticular nucleus and $I_{Th}$ a constant
tonic activity.  $IT_c^{in}$ is fed to the reinforcement learning
module for the color ($AC_{color}$). Details of the reinforcement
learning are given below (Section~\ref{sec:AC}).  Then, the resulting
channels along with $IT_c^{out}$ are given as inputs to the BG.  For
full details about Th and BG model see \citep{Girard2008}.


\begin{figure}
\centering
  \includegraphics[width=0.7\linewidth]{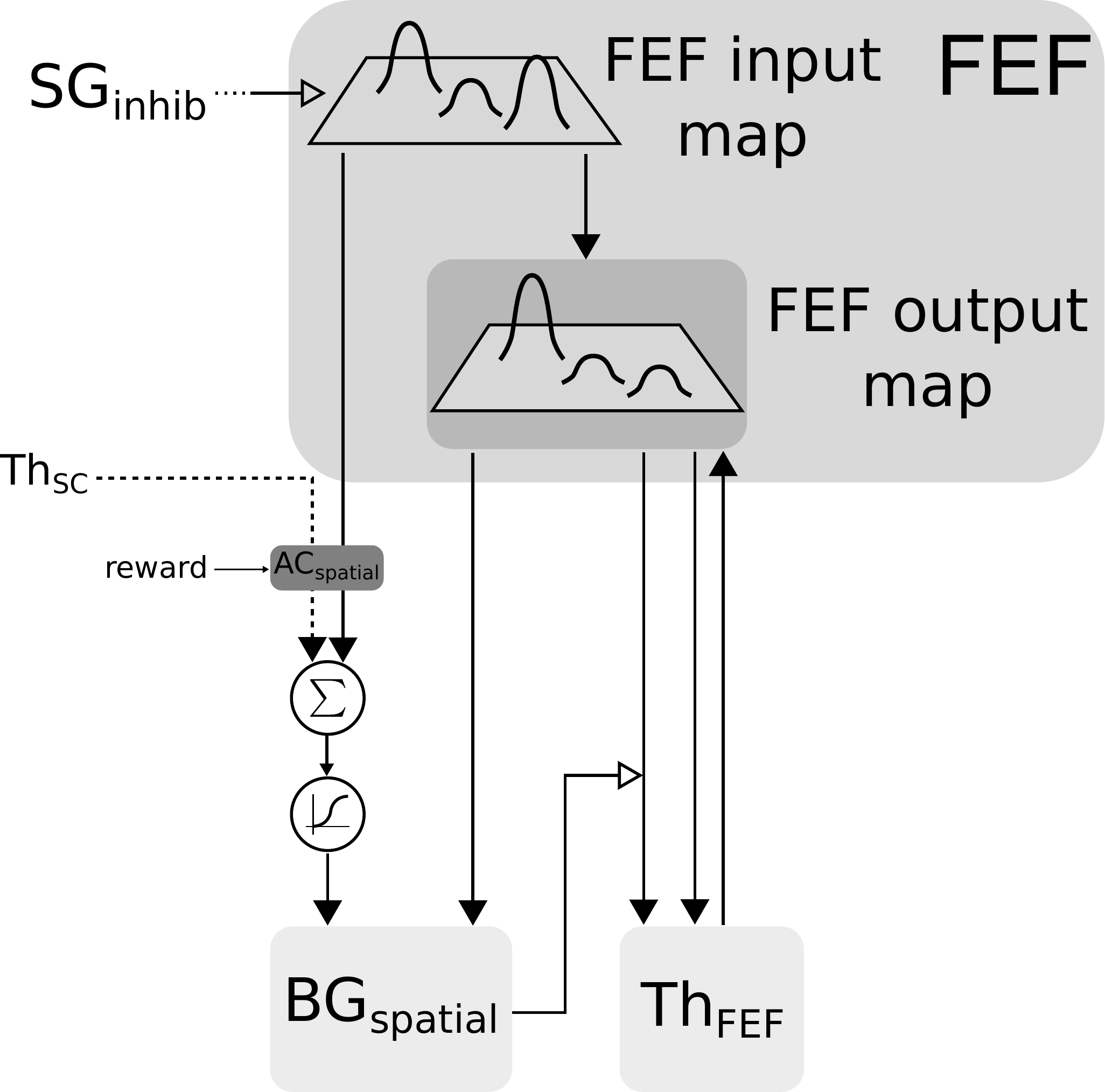}
  \caption{\label{FEF} Closed loop selection-amplification of spatial
    FEF map. Black arrow heads are exitatory connexions, empty arrow heads are
    inhibitory connexions.}
\end{figure}

Spatial information is processed by two cooperating loops.  In the
cortical FEF-BG-Th loop, FEF receives visual information in its input
map with a long latency (91~ms). This map
is then fed to the selection loop (cf. Fig.~\ref{FEF}) and the resulting
activity is computed as follows:

\begin{align}\label{eq:FEF_loop}
  FEF_{i,j}^{out} =&
  \W{FEF^{in}}{FEF^{out}}.FEF_{i,j}^{in}\times
  (1-W_{SG_{inhib}}.SG_{inhib})\notag\\
  &+\W{Th^{FEF}}{FEF^{out}}.Th_{i,j}^{FEF}\\
  Th_{i,j}^{FEF} =& FEF_{i,j}^{out}\times \Big(
    \W{FEF^{out}}{Th^{FEF}}+\W{GPi|SNr}{Th^{FEF}} \times (1-\frac{GPi|SNr_{i,j}^{spatial}}{GPi|SNr_{rest}})
    \Big) \notag\\
&-\W{TRN^{FEF}}{Th^{FEF}}.TRN^{FEF} \notag+I_{Th} \notag\\
\end{align}

with $(i,j) \in [0,n]^2$, $FEF^{in}$ the visual input and $FEF^{out}$ the
activity of the $FEF$ layer connected with $Th^{FEF}$.

The two maps ($Th_{SC}$ and $FEF_{in}$) are concatenated and fed to
the reinforcement learning module. We decided to keep both maps concatenated in
order to preserve the full learning capabilities and then to merge
back the resulting weighted maps at the $BG_{spatial}$ input level
before BG selection.

The merge is done by summing and passing these maps through a sigmoid
($f(x)=\frac{1}{1+e^{15.(0.95-x)}}$), inducing a non-linearity and a
minimal salience threshold. Similarly to the color loop, the resulting
map along with $FEF^{in}$ are given as inputs to the BG.


\begin{figure}
\centering
  \includegraphics[width=0.7\linewidth]{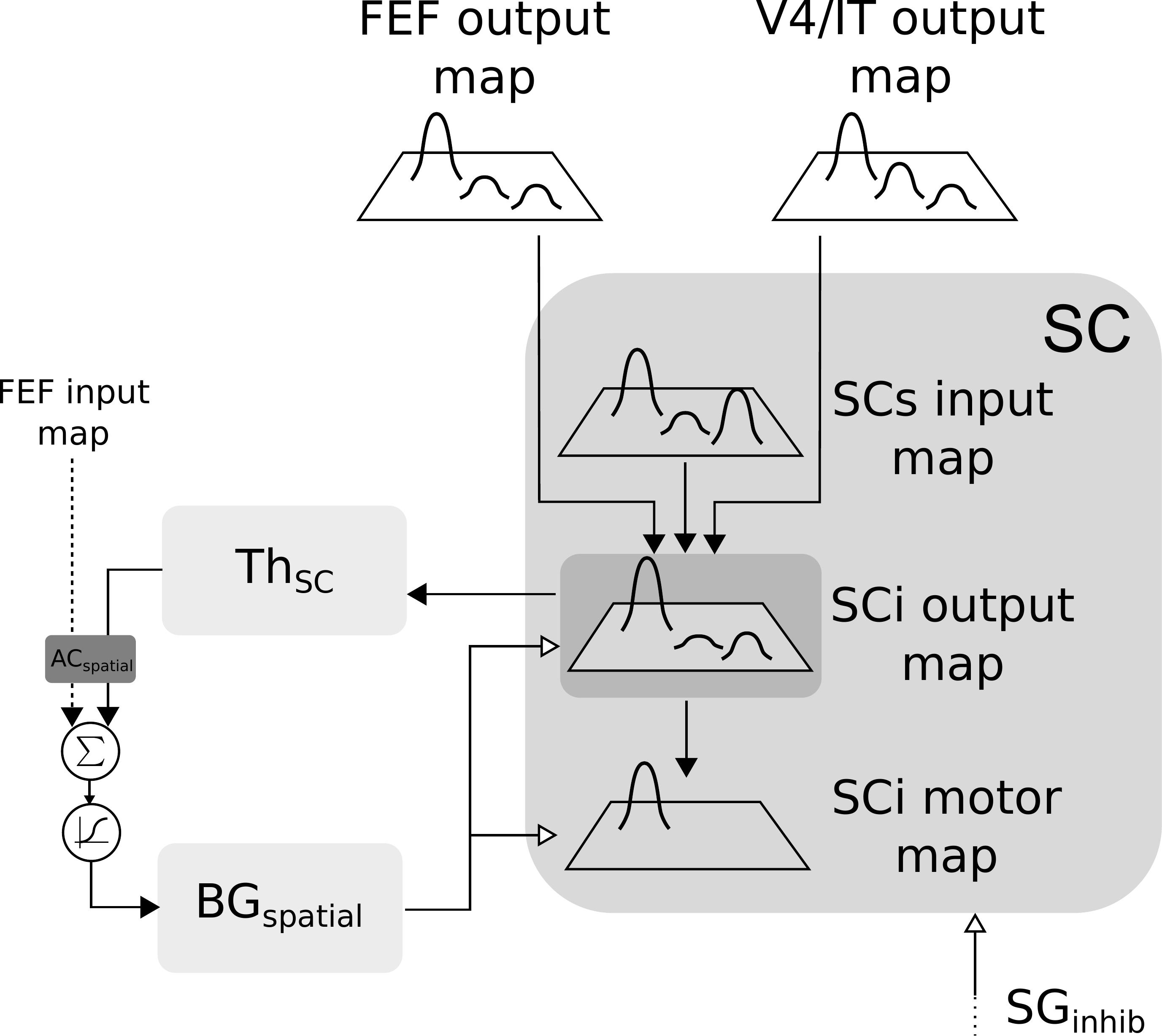}
  \caption{\label{fig:SC} Closed loop selection-amplification of spatial
    SC map. Black arrow heads are exitatory connexions, empty arrow heads are
    inhibitory connexions.}
\end{figure}

In the SC-Th-BG loop, SCi receives inputs from V4\textbar IT, FEF and retina
(via SCs). These inputs are weighted summed and fed to the selection loop
(cf. Fig.~\ref{fig:SC}). As stated previously the $BG_{spatial}$ module is the
same than in the FEF-BG-Th loop. The resulting activity is computed as follow:

\begin{align}\label{eq:SC_loop}
    SCi_{i,j}^{out} =& \big[ (\W{SCs}{SCi}.SCs_{i,j} + \W{FEF}{SCi}.FEF^{out}_{i,j} + \W{V4|IT}{SCi}.V4|IT^{out}_{i,j}) \big]\notag\\
   & \times \big[ \W{SCi^{in}}{SCi} + \W{BG_{amp}}{SCi} \times (1-\frac{SNr_{i,j}}{SNr_{rest}}) \big] \notag\\
    &\times (1-W_{SG_{inhib}}SG_{inhib}) \\
    Th_{i,j}^{SC}=&SCi_{i,j}^{out}\times \Big(
    \W{SCi^{out}}{Th^{SCi}}+\W{GPi|SNr}{Th^{SCi}} \times (1-\frac{GPi|SNr_{i,j}^{spatial}}{GPi|SNr_{rest}})
    \Big) \notag\\
&-\W{TRN^{SCi}}{Th^{SCi}}.TRN^{SCi} \notag+I_{Th} \notag\\
\end{align}

with $(i,j) \in [0,n]^2$, $SCs$ the visual input from the superficial
layer of SC, $GPi|SNr$ the inhibition from the output nucleus of BG
projecting to SC.


\subsubsection{Basal Ganglia}

The Basal Ganglia model used here was first described in
\citep{Girard2008} and is depicted in Figure~\ref{BG_all}A for
cortical loops.
Notice that for the subcortical loop the connectivity is slightly
different for the position of the Thalamus (cf. Fig.~\ref{BG_all}B). 

The parameters of the BG circuit involved in the spatial loop have
been adapted so as to cope with the selection of 630
channels (see Table~\ref{tab:BG_spatial_params}).

The $n \times n$ inputs from the spatial maps (here with $n=70$ for each
hemifields) converge on the $m \times m$ inputs (here $m=18$ for each
hemifield) by the Gaussian Pyramids method. Input map size is reduced
by first convolving it with a $5 \times 5$ gaussian kernel:

\begin{equation}
  \label{eq:Conv_in}
BG^{spatial}_{i,j}=(In*G)_{i,j}
\end{equation}
with $In$ the input map, $G$ the normalized gaussian kernel and
$(i,j) \in [0,n]^2$. Then it is $2\times 2$ binned in order to divide
dimensions by 2.

This operation is repeated 2 times in order to reduce
the input map by a factor 16.  The opposite operation is computed to
upscale the output activity of BG in order to match the projection
toward other structures.  This dimensionality reduction is inspired by
the anatomy of cortico-striatal connections \citep{Zheng2002}.

As seen on figure~\ref{featloc}, the color BG circuit receives the
sums of the activity of the color maps, and thus operates selection among three channels:

\begin{equation}
  \label{eq:V4BG}
  BG_c^{color} = \W{IT_c}{BG} \sum_{i,j} IT_{c_{i,j}}
\end{equation}
with $\W{IT_c}{BG}$ a normalization constant.
The output of the same circuit thus affects the whole color maps in
the following manner: 

\begin{equation}
  \label{eq:BGV4}
    V4^{out}_{c_{i,j}} = \widehat{V4}_{c_{i,j}}.IT^{out}_c  
\end{equation}
with $\widehat{V4}_{c_{i,j}}$ the normalized activity of the input map
for color $c$,
$\widehat{V4}_{c_{i,j}}=V4^{in}_{c_{i,j}}/max(V4^{in}_{c})$ and $IT^{out}_c$
the output activity for a whole channel $c$.

\textbf{Figure 7.~}{A: Details of the BG model in the cortical loop
    (here, IT-BG-Th is shown but an identical structure is used for
    FEF-BG-Th). Only 3 channels are represented, the middle one being
    the most salient. SNr/GPi and GPe are color inverted as channels
    activity in these structures are opposed (middle channel which is
    the most activated in input, is the weakest in these
    structures). Thalamus structure (Th) is composed of a ventral anterior
    nucleus and of reticular nucleus (TRN) which constitute a
    population without segregated channels. Striatum is composed of D1
    and D2 types of dopaminergic neurons and of a population of fast
    discharge inter-neurons (FS). Filled arrow heads are exitatory
    connexions and empty arrow heads are inhibitory. Filled lines
    represents one-to-one connexions and dotted lines represents
      one-to-all connexions. Adapted from \citep{Girard2008}.
B: Details of the BG model in the subcortical loop
    (SC-Th-BG). Same model than in A except for
    the position of the Thalamus.\figlabel{BG_all}}

\begin{figure}
\centering
  \includegraphics[width=1\linewidth]{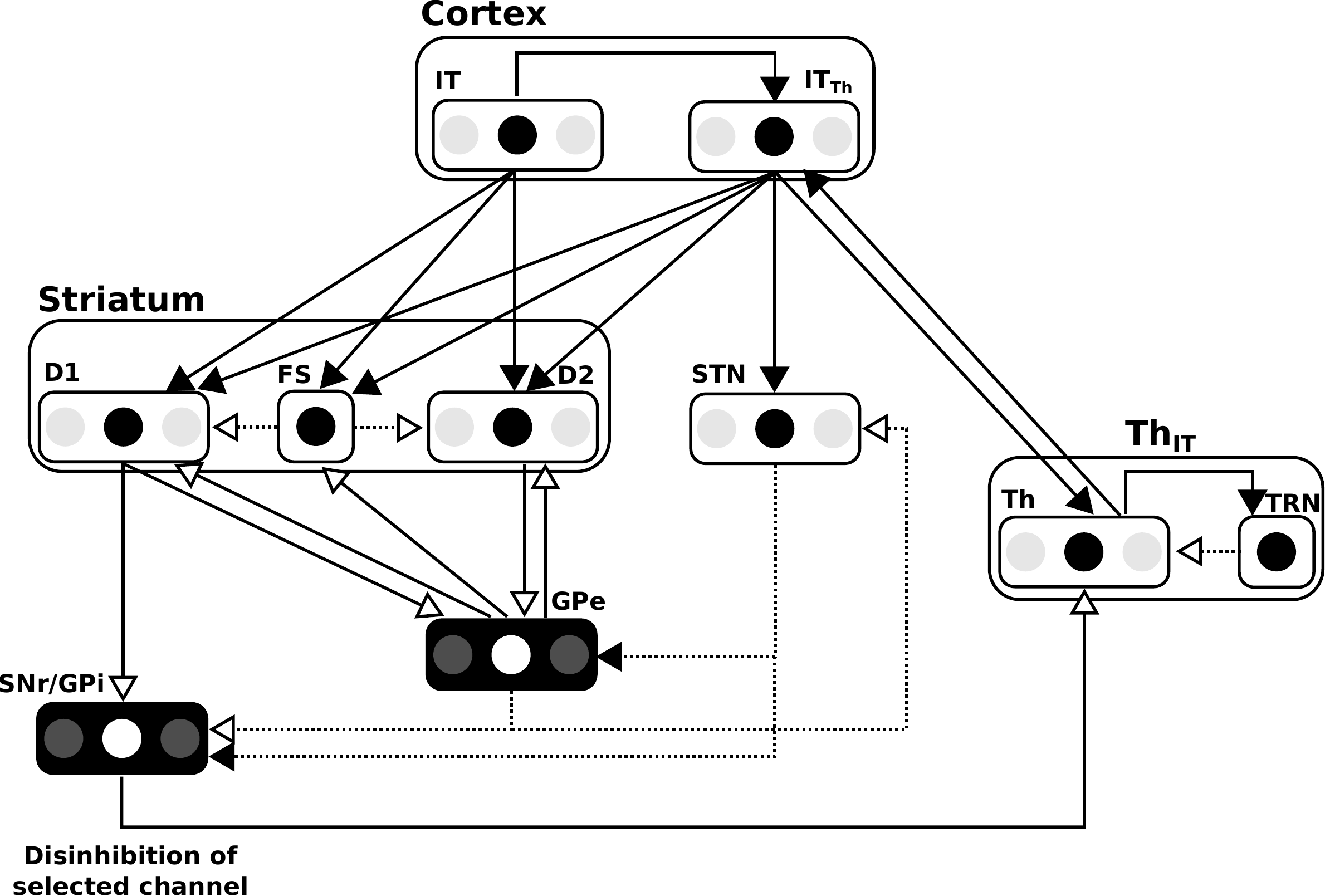}
  \caption{\label{BG_ctx}
A: Details of the BG model in the cortical loop
    (here, IT-BG-Th is shown but an identical structure is used for
    FEF-BG-Th). Only 3 channels are represented, the middle one being
    the most salient. SNr/GPi and GPe are color inverted as channels
    activity in these structures are opposed (middle channel which is
    the most activated in input, is the weakest in these
    structures). Thalamus structure (Th) is composed of a ventral anterior
    nucleus and of reticular nucleus (TRN) which constitute a
    population without segregated channels. Striatum is composed of D1
    and D2 types of dopaminergic neurons and of a population of fast
    discharge inter-neurons (FS). Filled arrow heads are exitatory
    connexions and empty arrow heads are inhibitory. Filled lines
    represents one-to-one connexions and dotted lines represents
      one-to-all connexions. Adapted from \citep{Girard2008}.
B: Details of the BG model in the subcortical loop
    (SC-Th-BG). Same model than in A except for
    the position of the Thalamus.}
\end{figure}


\subsubsection{Actor Critic \label{sec:AC}}

The input to the Basal Ganglia circuits is biased by reward using the classical
``Actor-Critic'' TD($\lambda$) learning algorithm \citep{Sutton1988,Sutton1998,Montague1996}.


\begin{figure}
\centering
  \includegraphics[width=0.8\linewidth]{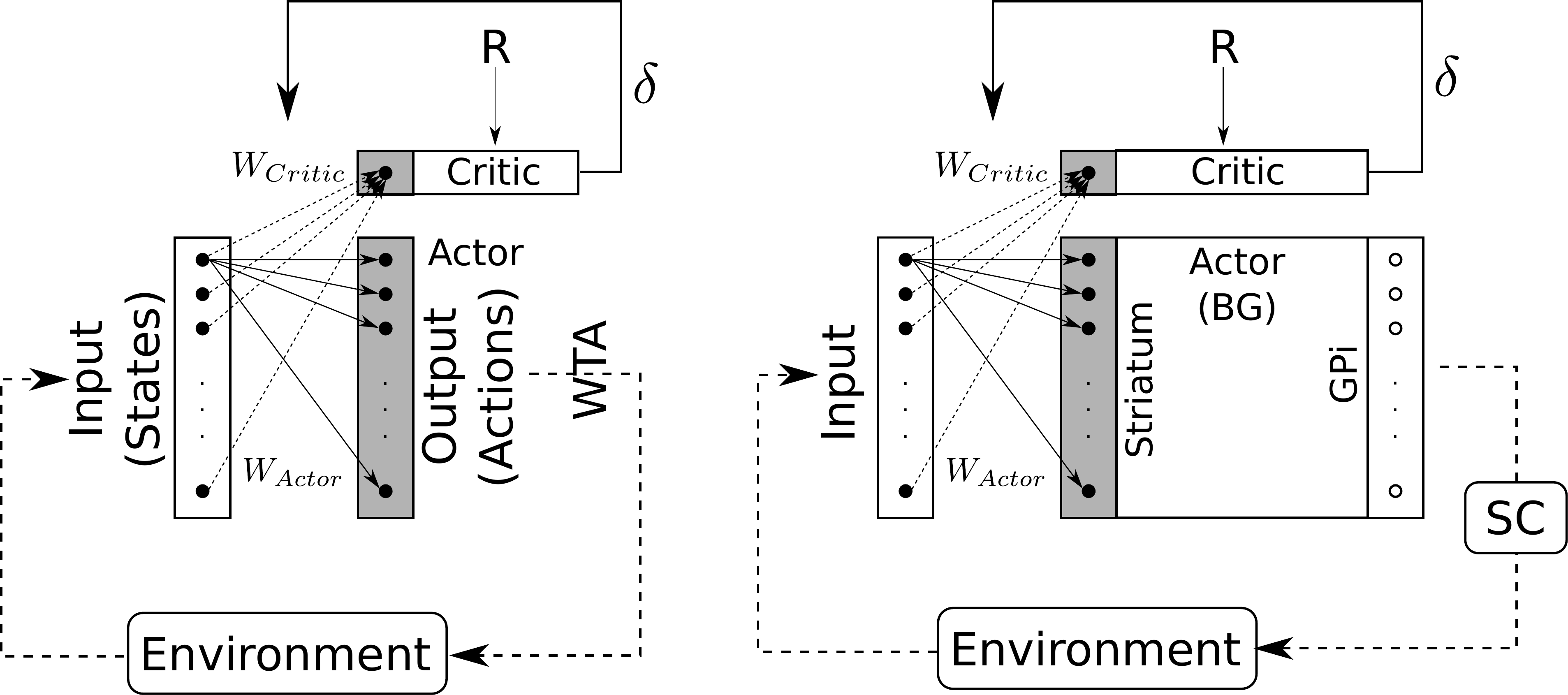}
  \caption{\label{fig:ac} Schematic representation of the Actor-Critic
    reinforcement learning algorithm. Left: classical Actor-Critic,
    involving a winner-takes-all (WTA) selection mechanism. Right:
    Actor-Critic with BG as selection mechanism.}
\end{figure}

TD-error $\delta$ is computed according to
\begin{equation}
  \label{eq:tderror}
  \begin{array}{c}
  \delta = R_t+(\gamma
  \times V_t) - V_{t-1}\\
  \mathrm{with} \\ 
V_t=W_{Critic} \cdot Input_t     
  \end{array}
\end{equation}
 
$R_t$ being the reward at time $t$, $V_t$ the estimated value
function, $W_{Critic}$ the learned weights of the Critic, $Input_t$
the input matrix (spatial or color) and $\gamma$ the discount factor.

Critic's weights are then updated using eligibility traces
$E_{Critic}$:
\begin{equation}
  \label{eq:wcritic}
  \begin{array}{c}
    W_{Critic} \leftarrow W_{Critic} + \eta \times \delta \times
  E_{Critic}\\
 \mathrm{with} \\
E_{Critic} \leftarrow \lambda \times E_{Critic} + Input_{t-1}
  \end{array}
\end{equation}

$\eta$ being the learning rate and $\lambda$ the ``forgetting'' factor
of eligibility traces. 
The size of the Critic's weights vector is $N$, the same as $Input$ so
here connexions are ``all-to-one'' type.
Actions vector (weighted inputs) is computed as following:
\begin{equation}
  \label{eq:actions}
A_t=W_{Actor} \cdot Input_t
\end{equation}
and Actor's weights are computed as following:
\begin{equation}
  \label{eq:wactor}
  \begin{array}{c}
W_{Actor} \leftarrow W_{Actor} + \eta \times \delta \times
E_{Actor}\\ 
\mathrm{with}\\ 
E_{Actor} \leftarrow \alpha \times E_{Actor} + Input_{t-1}\otimes
A'_{t-1}\\
\mathrm{and}\\ 
A'_{t-1}= GPi_{t-1}
  \end{array}
\end{equation}

Actor's weights matrix is of size $N\times N$ so here, connexion are
``all-to-all'' type.

Compared to classical reinforcement learning (cf. fig.~\ref{fig:ac},
left) we can see that ``States'' are inputs to be selected and
``Actions'' are weighted inputs. Here, the BG compute a selection of these
weighted inputs -- thus playing the role of the ``winner-takes-all''
(cf. fig.~\ref{fig:ac}, right) -- and then disinhibit some structure
(i.e. SC) which eventually will trigger a real action.

Actor's weights are initialized to an identity matrix in order to
allow for an initial ``standard'' behavior (direct unweighted projection). A minimum value for
Actor's weights diagonal has been implemented ($W_{Actor_{min}}=0.6$) in order
to prevent the system to from losing the ability to trigger saccades.
Critics weights are initialized to a random matrix with values $\in
[0,0.01]$.

The exploration, which is important for RL convergence , is caused
here by a perceptual noise only. This perceptual noise is implemented
in the following manner: one input has an amplitude of 1 and the other
of 0.95.
This 5\% difference is sufficient for the system to select the most
``intense'' input, before learning adds its own biases to the
selection, and is randomly alternated between cues in order to ensure
the absence of a systematic bias.

\subsubsection{Spatio-temporal transformation}

\begin{figure}
\centering
  \includegraphics[width=0.4\linewidth]{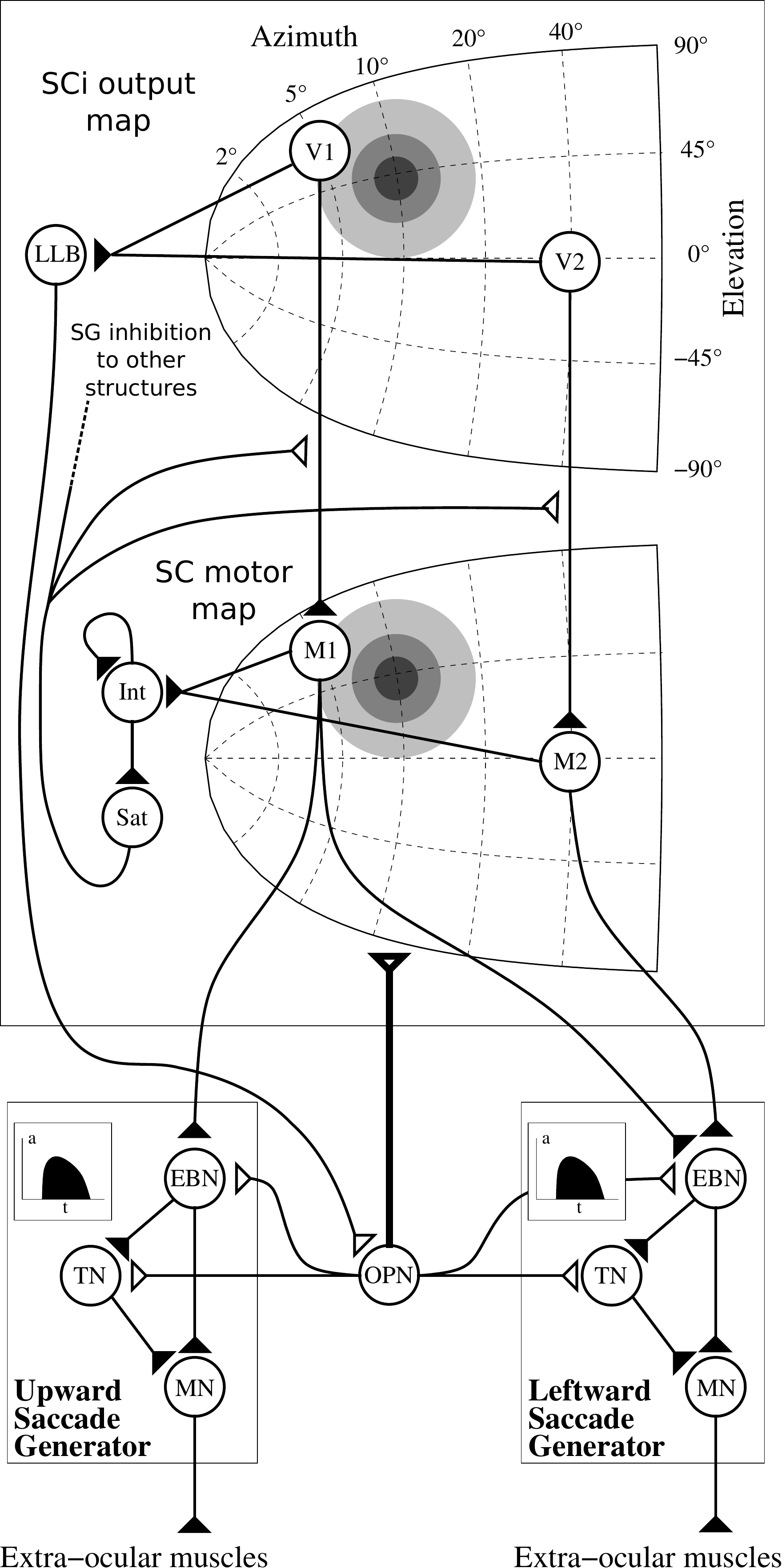}
  \caption{\label{SC_details}
 Architecture of the motor layer
    of SC. Only one colliculus (right hemifield) and two SBG are
    represented (without cross projections) along with two neurons by
    map (V1 and V2 in the SCi output map, M1 and M2 in the motor map).
    Grey discs represents gaussian activity produced by a visual
    target (coordinate ($10\,^{\circ}$,$10\,^{\circ}$), thus
    $R=10\,^{\circ}$, $\theta = 45\,^{\circ}$), insets in the saccade
    generator represent the temporal coding in EBNs generated to
    control muscles. Filled triangles are for excitatory connexions,
    empty triangles are for inhibitory connexions. Bold connexion
    affect the whole map. Adapted from \citep{Tabareau2007}.}
\end{figure}

In order to compute the so-called ``spatio-temporal transformation'' (STT)
required to convert a spatially coded target into a saccade burst
generators (SBGs) temporal sequence, we used the model first described
in \citep{Tabareau2007} (cf. Fig.~\ref{SC_details}).
This model includes a visual map (SCi output map described above) and a motor map (SC
motor map) with a log-complex mapping along with
colliculi gluing mechanism. The motor layer is projected to the saccade
generators and both are controlled by a strong inhibition from
omnipause neuron (OPN).


We can notice than we slightly modified the ``integrating-saturating''
mechanism ($Int$ and $Sat$ in figure \ref{SC_details}). This mechanism no
longer inhibits the whole motor map in a subtractive manner, but now
modulates the visual map to motor map projection in a multiplicative
manner:

\begin{equation}
  \label{eq:scinhib}
  \begin{array}{lll}
    I^{motor}_{i,j} &=& SCi_{i,j}^{out} \times
    (1-\W{BG_{inhib}}{SCi}\times SNr_{i,j})\\
    &&\times (1-\W{Sat}{Mot}.Sat)\\
    &&-\W{OPN}{Motor}.OPN
  \end{array}
\end{equation}
with $I^{motor}$ the input activity of motor layer, $SCi^{out}$
the activity of $SCi^{out}$ map described in section \ref{sec:loops}, $OPN$ the output activity of the OPN
and $(i,j) \in [0,n]^2$.

This modification has the advantage of generating more realistic burst
activities, more similar to the gamma functions used in \citep{vanOpstal2008}. 

Notice that $Sat$ is used as the ascending inhibitory signal
$SG_{inhib}$ in other structures, which signals the execution of a
saccade \citep{Sommer2002} .

\subsubsection{Model parameterization}

The parameters of the model were hand-tuned, these tuning operations
were performed, as much as possible, by considering the various
subsystems (BG models, generation of the motor command, convergence of
the inputs on the SC, and reinforcement learning) in isolation and
enforcing their correct operation.

The parameters of the spatial BG loop had to be modified compared to
the initial parameterization of (Girard et al., 2008), as the number
of competing channels is much higher. This drastically affects the
effects of diffuse projections, like those of the STN on the GPe and
GPi. When 630 channels are exciting the GPi, rather than 6, the
strength of this excitation has to be reduced, so as to avoid
saturating the GPi neurons, and so as to allow one-to-one inhibitions
from the Striatum to be strong enough to conteract excitation and thus
allow selection. These modifications were made as follows: the BG
model was isolated from the rest of the system, and provided with 2D
Gaussian inputs similar to those used in the tasks, with varied
amplitudes. The parameters were adjusted until the selection of a
single target with an amplitude between 0.6 and 1 was restored. Finer
adjustment were then made so that one or two distractors of inferior
amplitudes would not disturb the selection process, and that the
simultaneous selection of multiple targets occurred only when they
have very close amplitudes.

The parameters of the motor layers of the SC, and of the saccade
generators, which operate the spatio-temporal transformation, were
almost identical to those of \citep{Tabareau2007}, except slight
modifications in the integration rate of the saturating mechanism, so
as to adjust the duration of the motor bursts to more realistic
values.

The parameters adjusting the strength of the contributions of all the
different maps to the final $SCi$ layer were adjusted so that: 1)
imposing an input from the spatial system only, or from the color one
only, would generate the corresponding saccade, and 2) simultaneously
imposing a given target position in the spatial system and another one
in the color system, would result in an averaging saccade.

Finally, the parameters driving the temporal integration of reward in
the learning modules --namely the discount factors $\gamma$ and the
eligibility trace $\lambda$-- had to be large enough, so that learning
could occur despite the relatively long delay between the appearance
of a target and the effective reward delivery ($\approx500ms$). The
learning rates were adjusted so that the learning would converge to
the best possible level of performance in approximately $20-25$
sessions. The relative difference between $\eta_{spatial}$ and
$\eta_{color}$ has to be considered in the light of: 1) the huge
difference in the number of input weights to be adjusted in each
system ($1587600$ in the spatial domain vs. $9$ in the color one), and
2) the different extent of the input stimulations corresponding to one
target (a 2D Gaussian input spreading over a hunded of channels
in the spatial domain vs. one single channel in the color domain).

\subsection{Simulated tasks}


\begin{figure}
\centering
  \includegraphics[width=0.7\linewidth]{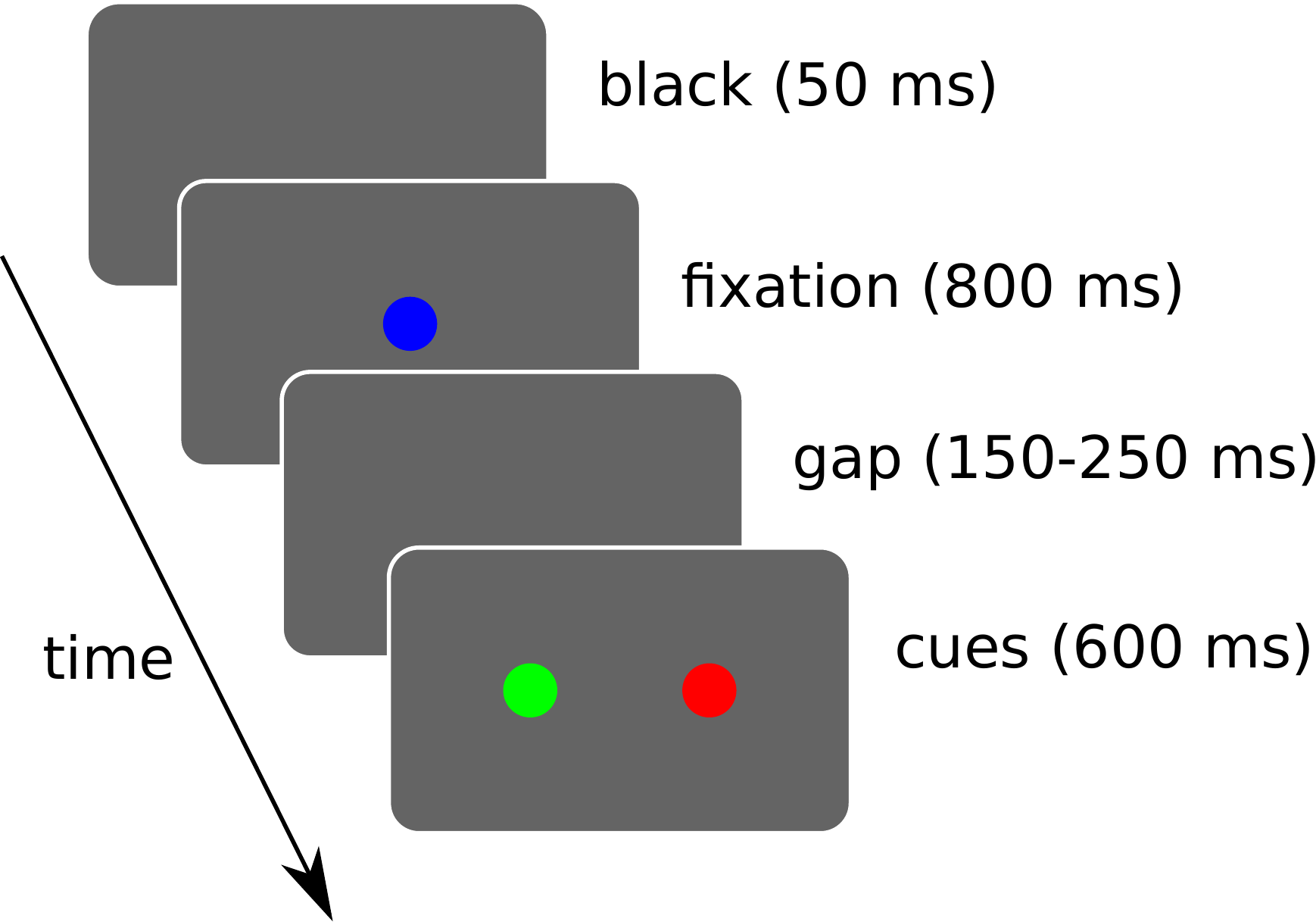}
  \caption{\label{fig:task}
Simulated sequence of visual stimuli. A
    black screen of 50~ms is followed by a fixation cue for
    800~ms. Then a random gap time (between 150 and 250~ms) is
    followed by the two cues. The cues are displayed for a maximum of
    600~ms and loops back. During this interval, if a saccade of
    sufficient amplitude ($>2.5\,^{\circ}$ from the center) is
    detected, the trial ends and loop back. Rewards are given when the
    trial ends, which may be triggered by the timer or a saccade
    depending on the task. }
\end{figure}

We simulated 3 target selection tasks where the system has to
trigger a saccade toward one of the two displayed cues
(cf. Fig.~\ref{fig:task}).

A ``spatial task'' is aimed at verifying
its ability to learn to choose a target based on spatial information only. A ``color
task'' for color information only. And a ``conjunction task'' to study
interactions between these two.   
10 runs were done and each experimental run is composed of 40 sessions of
12 trials.

\section{Results}

\subsection{Spatial task}

In the spatial task, the rewarded cue only depends on its position on
the visual field. So
the system has to learn to ignore the color information and to favor
the spatial one.


\begin{figure}
\centering
  \includegraphics[width=0.7\linewidth]{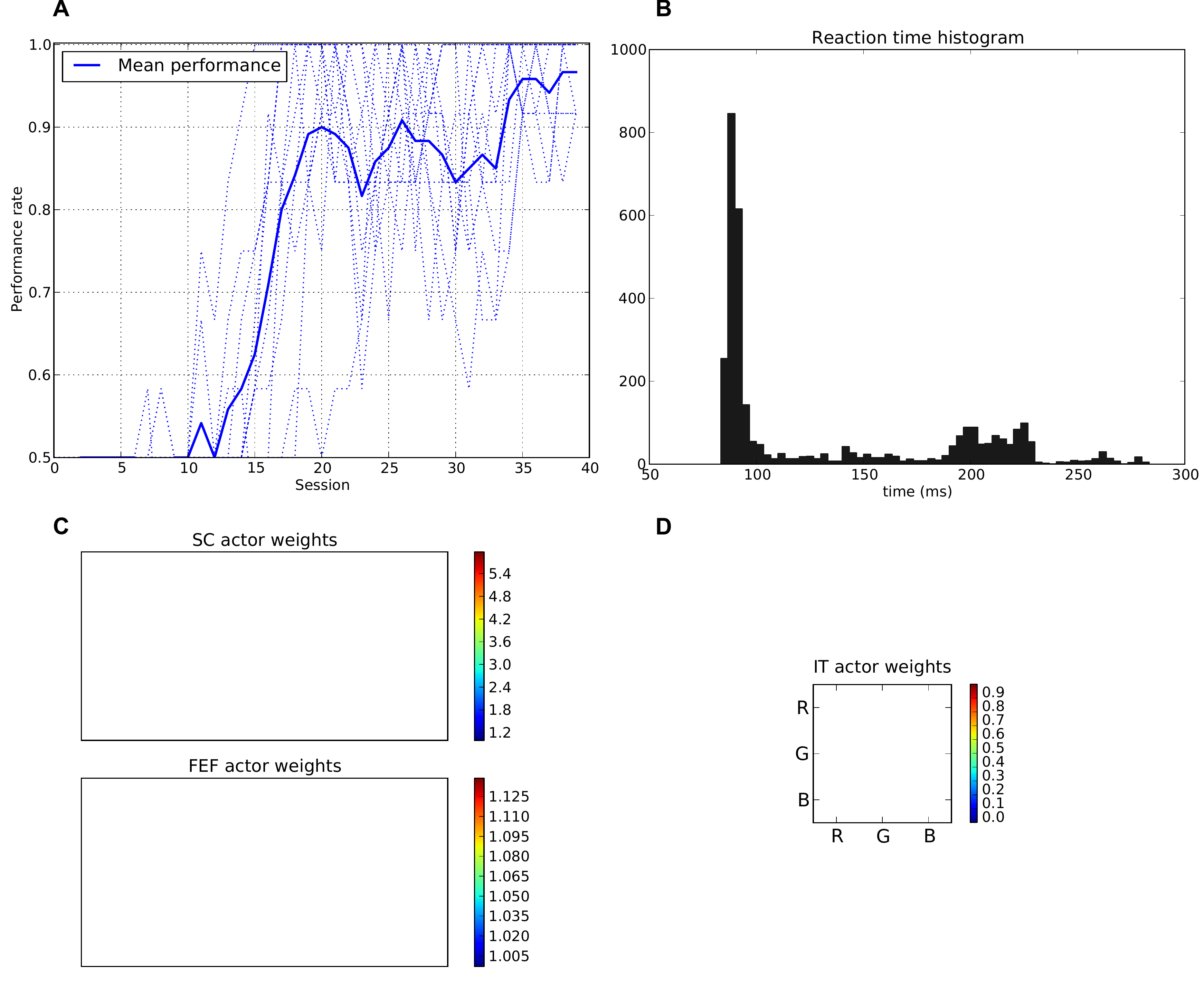}
  \caption{Results of the spatial task. The rewarded cue is the right
    one regardless its color. A: Performance across sessions (bold
    line is the mean performance of the 10 runs represented with
    dotted lines). B: Distribution of
    saccadic reaction time (SRT) for the whole experiment. C: Learned
    weights (averaged over 10 runs) for the Actor part of the spatial
    loop; for readability reasons, the multidimensional weight matrix
    has been projected on the output: it represents, for each unit,
    the sum of the input weights coming from the whole map, for the SC
    (top) and the FEF (bottom), note also the different intensity scale between
    SC and FEF. D: Learned weights (averaged over 10 runs) for Actor
    part of the color loop. \label{fig:spatialtask_res}}
\end{figure}

We can see that the model is able to learn the task with a performance
reaching $\approx 90-95\%$ (Fig.~\ref{fig:spatialtask_res}A), this
means that it is possible to find a parameterization of the model
allowing for a good level of performance after learning

\begin{figure}
\centering
  \includegraphics[width=0.7\linewidth]{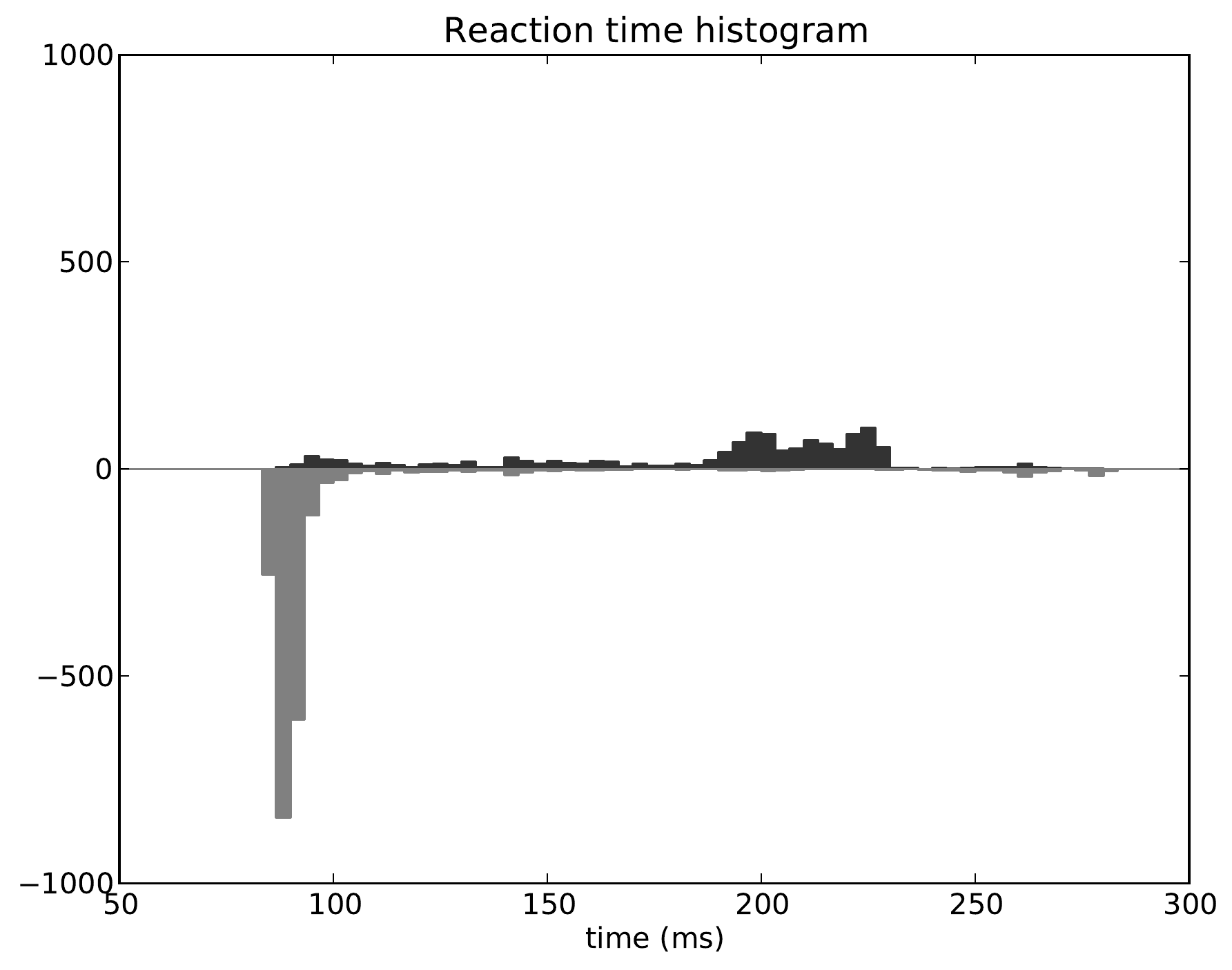}
  \caption{Spatial task reaction time histogram with separated first
    half of the experiment (top) and second half (bottom).\label{fig:spatialtask_rt_cut}}
\end{figure}

The distribution of SRT is bimodal, with a very sharp peak of low
latency ($\approx 88$~ms) and a second bump centered
around $\approx 200$~ms (cf. Fig.~\ref{fig:spatialtask_res}B).  This
behavior is very similar to that of ``express saccades'' for short
latencies and ``regular saccades'' for longer ones described in
\citep{Fischer1993}.  Looking at details of the evolution of these
SRT, it appears that for the first half of the experiment (first 20
sessions = first 240 trials) saccade latencies mainly fall within the
200~ms mode (cf. Fig.~\ref{fig:spatialtask_rt_cut}). These saccades reflect the
baseline timings of the system without any selection bias from learning. 

For the second half of the experiment (where performance is close to 90\%), saccade
latencies fall within the 88~ms mode.

Associated weights for the color loop
(Fig.~\ref{fig:spatialtask_res}D) indicates that the colors of targets
(red or green) have not been learned: they have similar
weights values of $\approx 0.8$ in the diagonal.

In contrast, the weights of the spatial loop
(Fig.~\ref{fig:spatialtask_res}C) show a strong bias toward the right
target (the rewarded one), especially in the weight map corresponding
to the SC ($\approx 5.5$, while the FEF ones are around $1.1$).  These
weights causes a strong activity on the spatial loop with a quick
disinhibition from the SNr as soon as the direct retina-to-SC signal
appears. Then, activity is transmitted to the motor layer even before
visual information reaches the cortical visual areas and rapidly
triggers a saccade. This kind of saccade thus differs from
``standard'' ones as they only rely on the direct retina-to-SC
pathway.  Indeed, before learning, the retina-to-SC input is not
sufficient to trigger a saccade alone in our model and needs either
FEF or V4\textbar IT input, thus explaining the longer SRT.


\begin{figure}
\centering
  \includegraphics[width=0.7\linewidth]{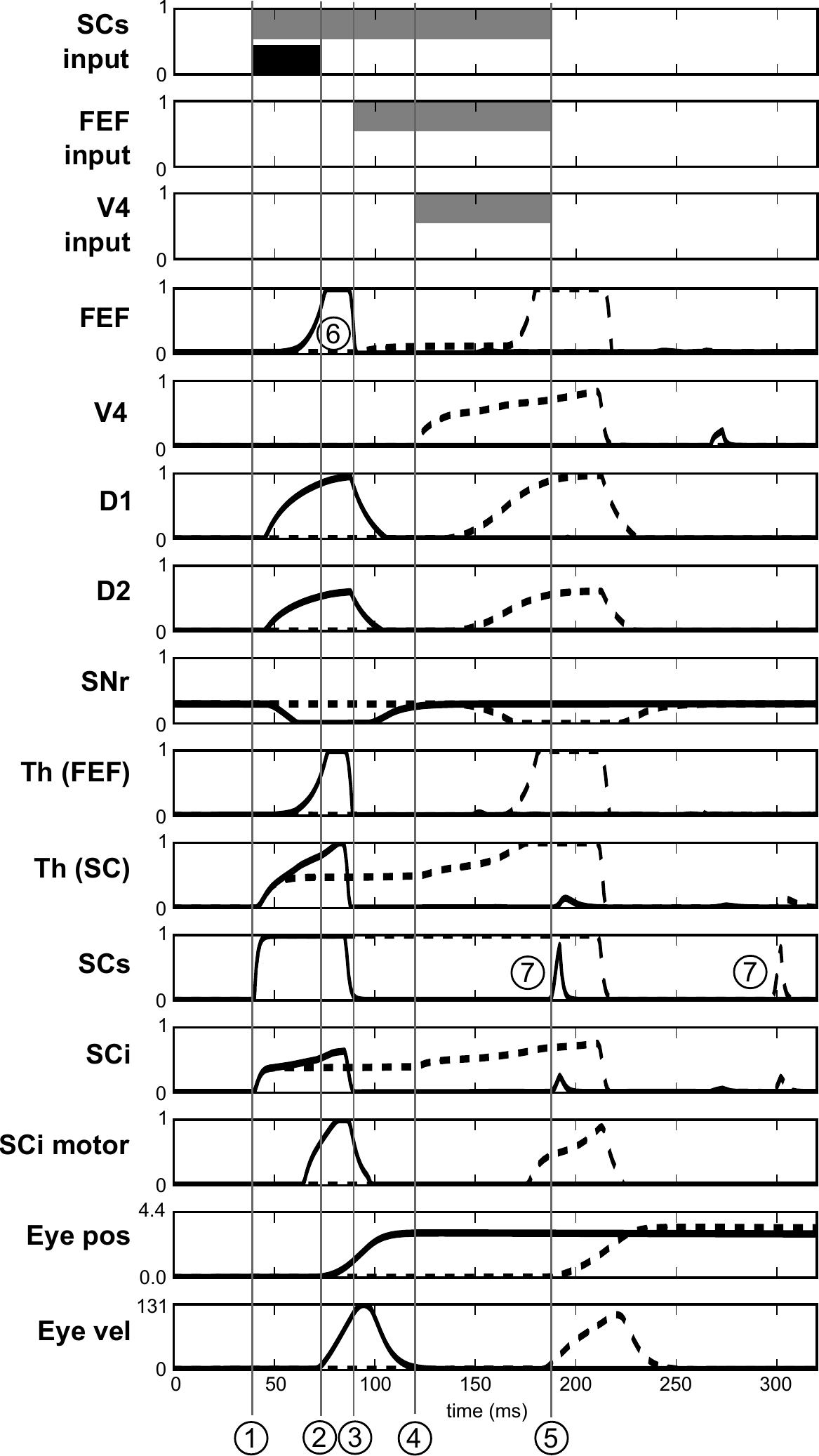}
  \caption{\label{allactivities}
Activities of different neurons of
    the target channel in the spatial task. Target appears at $t=0$. Dashed line: before
    leaning. Solid line: after learning.
    SCs input, FEF input and V4 input represent the presence of the
    visual cue in the receptive field before learning (gray) and
    after learning (black). 
    1: Visual activity reaching
    SCs. 2: Beginning of the express saccade (after learning). 3: Visual activity reaching FEF. 4: Visual activity
    reaching V4\textbar IT. 5: Beginning of the saccade before learning. 6:
    Indirect short latency activity in FEF provoked by SC activity. 7:
    Small burst of post-saccadic visual activity provoked by the end of
    inhibition from $SG_{inhib}$. }
  
\end{figure}

If we look at the details of neural activity in normal and express
saccades (Figure~\ref{allactivities}), what appears
for the spatial task (after learning) is that direct retinal input
induces activity in the spatial loop, which is quickly dis-inhibited
by the BG (thanks to the strong weights) and activates the SCi motor
map.  Moreover, as the same BG module is shared between the
subcortical and the cortical loops, this dis-inhibition also affects
the cortical loop and thus induces activity in FEF before visual
information reaches it.  This induced activity depends in facts on the
baseline level of the Thalamus and is a prediction of
the model due to our choice of a single shared spatial BG module.
The activity in the SC causes a disinhibition in the spatial BG
circuit, which then disinhibits also the thalamo-FEF loop. As this
loop is auto-excitatory and as the thalamus has a baseline activity,
this trigger a
resonance between Cortex and Thalamus. Thus the observed short latency
activity in FEF is not caused directly by visual input but indirectly
by subcortical visual activity. 

Yet, express saccades depend only on the SC loop and FEF only has a
marginal impact on it. Nevertheless, simulations with a FEF inactivation (after
learning) extends SRT of $\approx 15~ms$, this FEF resonant activity
thus contributes to the global behavior.

Notice that Figure~\ref{allactivities} also exhibits some very short
bursts of post-saccadic visual activity (better seen for SCs but the
mechanism is the same for all the structures). These bursts are
provoked by the residual retinal activity reaching each visual region
due to the latencies, whereas eyes have already moved. This behavior
is probably not significant as it may be canceled by a different
choice of parameters for $SG_{inhib}$ for example.


\subsection{Color task}


\begin{figure}
\centering
  \includegraphics[width=0.7\linewidth]{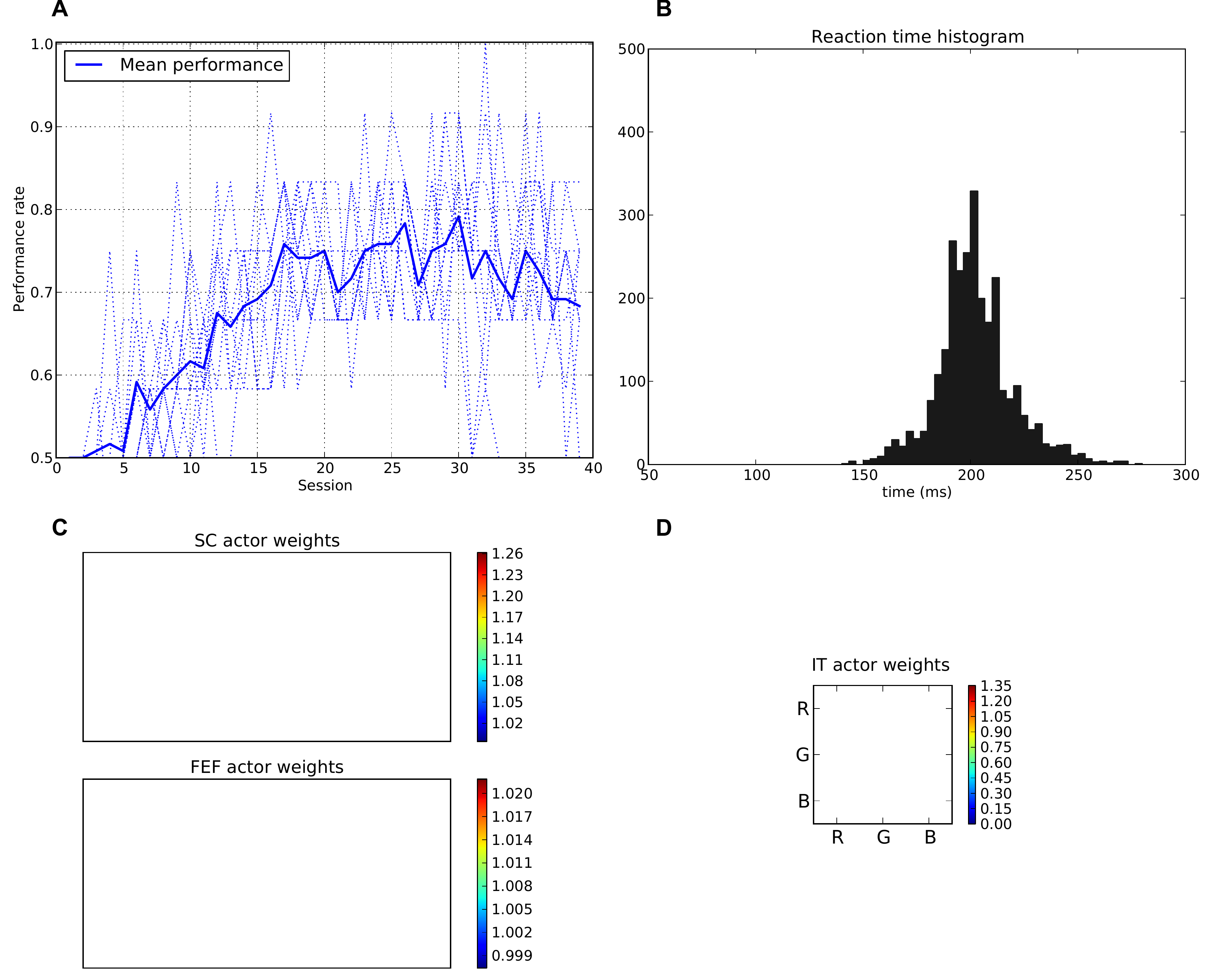}
  \caption{Results of the color task. The rewarded cue is the red one
    regardless its position. A,B,C,D: same as Fig.~\ref{fig:spatialtask_res}.\label{fig:colortask_res}}
\end{figure}

In the color task, the rewarded cue only depends on its color. So
the system has to learn to ignore the spatial information and to favor
the color one.

Here, the average performance only reaches about 75\%
(cf. Fig.~\ref{fig:colortask_res}A), so the system can learn the task
but errors are still made at a rather consistent rate. The performance
is thus lower that in the spatial task, an effect which is most probably
caused by the structure of the BG loops themselves, a point we discuss
further in section~\ref{sec:discudominspat}.

Color learning is very sensitive to noise in the spatial
domain. Indeed, most of the time ($\approx 95\%$) these errors occur
when the distractor (object with the wrong color) is the most intense
(``intensity'' is imposed to be 1.0 or 0.95 by the perceptual noise).
It means that even with learned weights favoring the good color in
average (cf. Fig.~\ref{fig:colortask_res}D) and spatial ones almost
symmetric (cf. Fig.~\ref{fig:colortask_res}C, the intensity scale
indicates very small variations), the color loop is sometimes unable to
impose its choice when a competition occurs between the spatial and
the color loop. This is explained by the fact that the subcortical
circuit, which operates exclusively on spatial information, can take
decisions faster than the color loop. It thus can impose a choice
based on spatial information even before the cortical color loop
converges to a decision.

The SRT for this task mainly consists on a single mode histogram
centered around 200ms, easily explained by the longer latency of the
color loop (122ms). No reduction of these latencies by learning were
to be expected, as no faster pathway operating on colors is available.

\subsection{Conjunction task}


\begin{figure}
\centering
\includegraphics[width=0.7\linewidth]{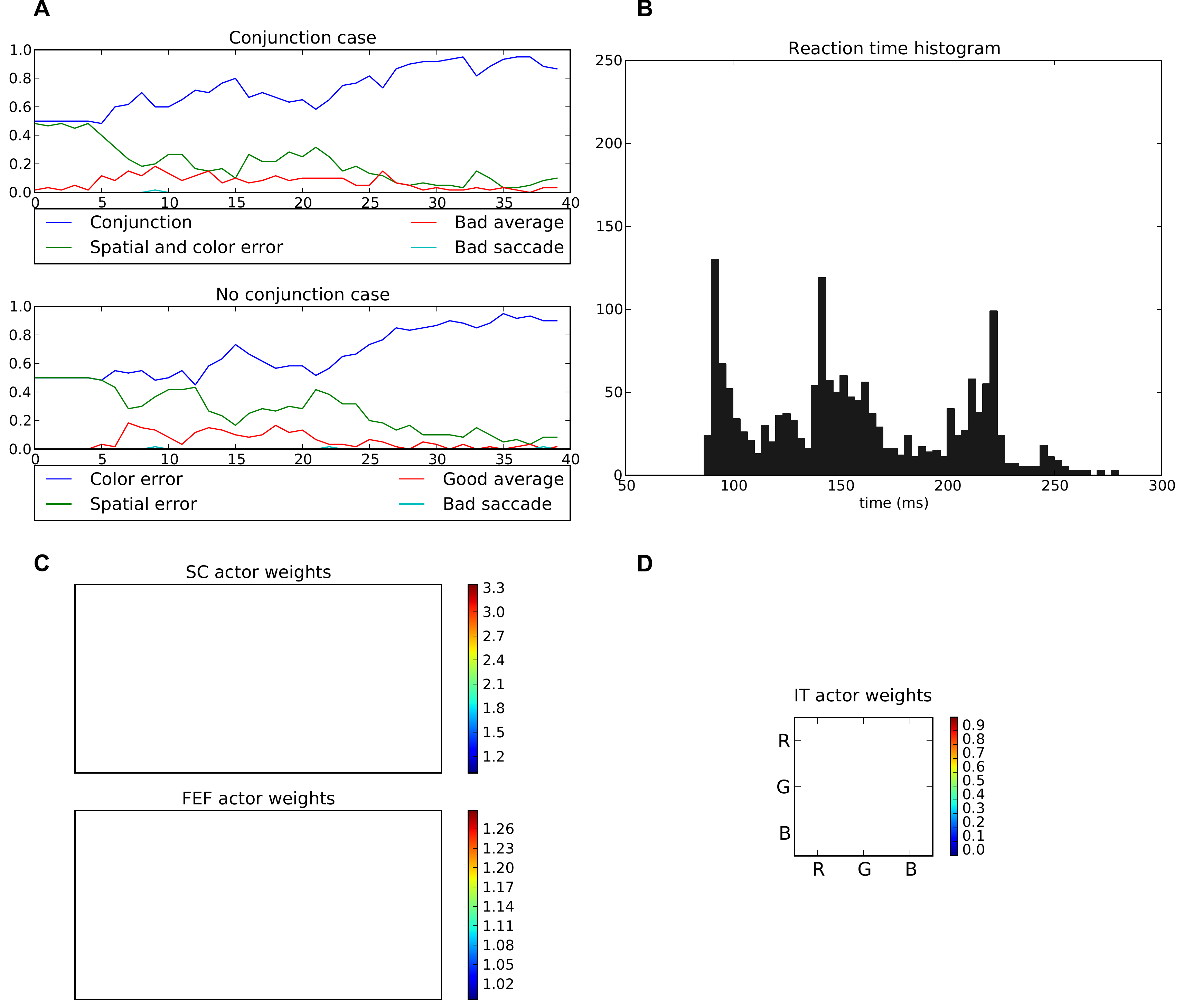}
\caption{
Results of the conjunction task. The rewarded cue is the
    right red one; if not present, reward is given for fixating the
    center area. A: average choices, in the conjonction (top) and no
    conjunction (bottom) case. In the conjunction case,
    ``conjunction'' represents the good choice, ``spatial and color
    error'' a movement towards the wrong cue, ``bad average'' an
    averaging saccade (both targets selected simultaneously) and ``bad
    saccade'' (saccades that fall neither within a 2.5~° radius from
    the center or any cue). In the no conjunction case, ``good
    average'' is a rewarded saccade keeping the eyes on the
    fixation point, ``spatial'' and ``color'' errors respectively
    represent movements to the green target on the right and to the
    red target on the left, and ``bad saccade'' in any other position
    (generally between fixation and cue but outside the 2.5~° radius).
    B,C,D: same as Figs.~\ref{fig:spatialtask_res} and
    \ref{fig:colortask_res}. 
    \label{spatialcolortask_res}}
\end{figure}

In the conjunction task, the rewarded cue depends on both position and
color (e.g. red disk at the right position). When this conjunction is
not presented (No conjunction case), the system is rewarded only if the eye
position stays within a 2.5° degrees circle around the center (``Good
average'' behavior). 

Here, the average performance for the conjunction case reaches levels
similar to those of the spatial task (around 95\%, Fig
~\ref{spatialcolortask_res}A) but for the ``No conjunction case'' the
rewarded behavior (``Good average'') is rarely performed.  We can see
that the errors made in this case tend to be mostly ``color errors''
i.e. a saccade toward the good location but with the wrong color
(around 90\% of errors at the end of the experiment). ``Spatial
errors'' occurred when a saccade is triggered toward the good color
but at the wrong position.  However, we can see that at the beginning
of the learning and until half of the experiment, the system is still
able to produce a small number of ``good average'' ($\approx
15\%$). This behavior progressively disappears as the spatial loops
learn and become faster, thanks to its subcortical component, making it
more difficult for the color loop to select due to its longer latency.

What appears at the end of the experiment is that the behavior of the
system is mainly dominated by the spatial loop with almost no
weighting from the color loop.

The learned weights correspond well to the task as the
right position is favored compared to left one with
(cf. Fig.~\ref{spatialcolortask_res}C) but the red color is only
slightly favored compared to green (cf. Fig.~\ref{spatialcolortask_res}D).

The saccade reaction time is more complicated here. In fact we can see
three modes ($\approx 88ms$, $\approx 140~ms$ and $\approx 220~ms$).
These three modes are in fact explained by the respective latencies
imposed for the three pathways, SC (41~ms), FEF (91~ms) and V4\textbar
IT (122~ms).  Similarly to the spatial task, most of the 88ms saccades
occurred on the second half of the experiment reflecting the
specialization toward spatial selection.  In fact, saccade latencies
shift from the 220~ms mode roughly at the first tier of the
experiment, to the 140~ms mode at the second tier and then to the
88~ms mode. This gradual shift of timing thus explain the lack of
influence of the color loop, whose pathway latency is of 122~ms. A
saccade may be triggered by the spatial loops before feature
information even reaches the color loop. Again, this effect is not
specific to a given parameterization: the advantage of the spatial
decisions, caused by a subcortical circuit with earlier access to
information, and thus with faster learning, is structural.  It is to
be noted that the 140~ms peak did not appear in the spatial task, as
the learning is fast enough to allow the system to quickly switch to
an ``express saccade expert''. This is also explained by the
information ``redundancy'' in our model between SC and FEF, the latter
dealing with the same spatial information only with a longer latency.
In the conjunction task, this peak appears as the ``difficulty'' slows
down the learning, and thus the shift to an ``express saccade
expert''.

\section{Discussion}

We described a model of the saccadic system with some very specific
structural features:
\begin{itemize}
  \item the cortico-basal circuits operate in various dimensions
    (selection based on spatial position, or on target features), with
    sensory inputs provided with a given latency,
  \item the subcortico-basal circuit operates on spatial
    information only, and with a shorter latency,
  \item all these circuits are subject to reinforcement learning at
    the level of the input of the basal ganglia,
\end{itemize}

We claim that this structure predicts very specific behaviors,
especially in feature-based and space-and-feature-based decisions:
\begin{itemize}
  
\item In the spatial decision task, an ability to switch, from
  long-latency to short latency saccades (thanks to the learning of the
  subcortical circuit). An effect experimentally described in \citep{Fischer1984}.

\item    In this task, after the learning of the subcortical shortcut, an
    early burst of activity in the FEF appears, caused by resonant
    activity in the spatial circuit. This burst slightly contributes
    to the reduction of the saccade latency.

\item    In the color decision task, the concurrently learning subcortical
    circuit reduces the efficiency of learning, when compared to the
    spatial task. In normal animals, this effect could be cancelled by
    a external cognitive brake, for example the dlPFC, acting on the
    subcortical circuit. Thus we predict that this deficit observed in
    simulation should be observed only in animals with prefrontal
    cortex deactivation.

\item    In the conjunction color-and-space-based task, again with the same
    prefrontal cortex deactivation, space should dominate in the sense
    that when the cunjunction is not presented, 1) inhibiting the
    response should be difficult and disappear with learning, 2) the
    resulting errors should be preferentially directed towards the
    correct position in space rather than towards the target with the
    correct color, 3) the saccade latencies should decrease as in the
    purely spatial task, a clear clue that the subcortical spatial
    circuit has taken full control of the decisions.
\end{itemize}

\subsection{Previous models}

Very few models have investigated the operation of multiple basal
ganglia circuits in saccadic decision and learning \citep{Girard2005},
and even fewer took into account the existence of a purely subcortical
loop.

The seminal model of Dominey \& Arbib
\citep{Dominey1992,Dominey1995} is quite complete, with memory and
sequence learning that we have not yet replicated.
Nevertheless, some of its aspects seems now rather
outdated.
First, their model lacks the subcortical SC-Th-BG loop which is now
clearly identified: they only integrated cortical loops. This
subcortical loop can operate faster than the cortical circuit and one
aim of our work is to explore their interactions.  
Second, the basal ganglia model they used is
oversimplified. Indeed it is only based on the direct/indirect interpretation of
the BG connectivity, from which they keep the direct
pathway only.
Consequently, concurrent channels cannot interact in the BG circuitry which make
target selection problematic. 
Their SC motor layer thus requires an ad hoc winner-takes-all
mechanism, where our more complete BG model solves these problems.

The model proposed in \citep{Brown2004}
includes a cortical loop dedicated to saccade strategy selection, and a
subcortical loop dedicated to target selection.
They also include a working memory mechanisms we have not yet
included.
Their cortical ``strategy'' loop explicitly selects whether the target of a saccade will be
based on the fixation cue, target position or target feature.
Their subcortical loop lacks any thalamic relay and is entirely
controlled by the cortical
loop, making it unable to learn and make saccade without it.
Finally, the details of their BG circuitry suffer from
limitations, discussed in details in \citep{Girard2005}.

\cite{Chambers2005} proposed a model integrating both the subcortical
and cortical pathways without learning capabilities, where a single
up-to-date BG model dedicated to location-based selection integrates
FEF and SC inputs. Using the various positive feedback loops of this
circuitry, they show that manipulating the level of dopamine in their
BG model generate reaction time and saccade size modifications
reminiscent of Parkinson's disease patient behavior.  
This model is equivalent to our spatial circuits, and does not explore
learning and competition between cortical loops.

The model described in \citep{Guthrie2013} integrates two cortical
loops (``cognitive'' and ``motor'') interacting through different
associative structures at both cortical and striatal level.
They store in a sub-part of the Striatum all the possible spatial and
feature combinations, which could create an obvious combinatorial
problem in a realistic model with a full field of view representation
and a rich feature space.
This model has shown the ability to learn to select targets based on
conjunction of information between the two loops but does not include
SC and does not specify how the selection in the BG is transformed in
a motor command.
The associative striatal structure is dependent on the associative
cortical one and provides a mean of information transfer between loops.
However, the BG architecture used is quite simplified, lacking GPe and
GPe-STN connectivity.
Finally this model does not include any subcortical loop and thus did
not study possible interactions between cortical and subcortical loops.

\subsection{Spatial dominance}\label{sec:discudominspat}

Our results show that the system is able to learn basic behaviors such
as the ``spatial task'' and the ``color task''. Moreover, we observed
quite different abilities for these tasks.  A first difference
appeared on the color task performance which only rises to about
75\%. This difference can be explained by the very structure of the
model where the spatial loop intrinsically dominates the system as it
includes the SCi output map and has access to information before the
color one. Thus, it can learn before the color loop processes
information and, has the last word on selection.  This characteristic
is confirmed in the ``conjunction task'' where the system finally
learned a ``spatial task''.  What is quite clear with this
architecture is that subcortical spatial choice should prevail when
opposed to a color one.  This characteristic was also observed in a
previous work with a simpler model without the cortical spatial loop
\citep{N'Guyen2010} and seems to be a prediction of this architecture.
Such a prediction could be tested on animals with dlPFC inactivation
in a task where both a spatial and feature criterion contradict each
other as we expect the dlPFC to inhibit impulsive subcortical
behavior. This prediction wouldn't be hold for the model proposed by
\cite{Guthrie2013} as they explicitly represent conjunction
information in the Striatum, and this allows for an experimental
discrimination between the two models.

\subsection{Express saccades}

Moreover, another stable outcome of this model relates to the saccade
reaction time. We observed what resembles to ``express saccades''
for the spatial task. These short latency saccades occurred only after
a period of learning in our case. This training dependent behavior is in
accordance with previous observations on monkeys (and humans)
\citep{Fischer1984, Fischer1986}. However it appears that monkeys are also able to
trigger some rare and spontaneous express saccades without learning
that our model cannot reproduce. This behavior may be viewed as a kind
of exploratory one, clearly lacking in our model.

These express saccades are only performed toward learned locations and
never toward learned features. This suggests that this behavior
is location dependent and not feature dependent, which is in accordance
with results in monkeys \citep{Fischer1984, Schiller2005}.  Indeed,
imposed sensory pathways latencies exclude the ability of express
saccade for the cortical color loop (122~ms) which easily explains the
lack of such saccade in the color task.  Therefore, the intrinsic architecture of
the model predicts that correct express saccades cannot occur based on
feature information.  Moreover in our system this spatial dependency is
encoded in a retinocentric reference frame and so doesn't depend on the
location of target in space which is also in accordance with previous
results \citep{Schiller2005}.

Moreover it seems that these express saccades are not dependent on FEF
as simulations done with FEF inactivation on a learned system, only
lengthen them of about 15~ms which seems to be quite in accordance
with what was observed in lesion studies \citep{Schiller1987b}.

Interestingly, we observed a short latency burst of activity in FEF
prior to the execution of the express saccade. This activity is not
caused by a direct visual input (it appears before visual input
reaches FEF) but by an indirect SC activity causing the a resonating
activity in the cortical loop.
Although a SC to FEF projection, either
direct or through the Thalamus, has been hypothesized
\citep{Sommer1998, Everling2000}, this induced activity through BG
disinhibition seems to be a new prediction of our model.  

Notice that the express saccades we obtained could be theoretically
shortened even more with a pre-disinhibition of BG which could be
viewed as a preparatory activity. Doing so it should be possible to
shorten latency by tens of milliseconds maybe explaining the observed
range of timings from 70 to 90ms in living animals.
For example a preparatory activity in FEF during the
gap period which could either facilitate or even elicit disinhibition
of BG \citep{Everling2000}. 
Whether this pre-disinhibition exists or not remains a question to be
answered experimentally.  
However this phenomenon was not observed in our system and may require
some memory capacity that we did not implement.  

If we look further at the SRT distributions, what is commonly observed
in primates is a bimodal distribution of reaction time for a detection
task (only one cue) which can be related to our spatial task. These
two modes are in the range of ~80-100ms and ~130-160ms.
Moreover, as
said before these timings keeps quite unmodified after a FEF lesion
but are drastically changed after a SC lesion \citep{Schiller1987b}.
Our model produces a compatible bimodal distribution but with a longer
latency for the second mode which involves the color loop. So it seems
that our model doesn't capture the exact mechanism explaining this
precise timing.

In contrast, a unimodal distribution is observed in primates for a
discrimination task (where the animal has to chose a cue based on a
feature) which can be related to our color task. In this case the
distribution is wider and in the range of ~160-200ms without express
saccades. Once again, this distribution remains unchanged after FEF
lesion but is modified after a SC lesion \citep{Schiller1987b}.
Here the mechanism proposed by our model seems quite consistent with
the experimental data.

Unfortunately to the best of our knowledge there is no data on a
spatial-feature conjunction task in the literature, but it is to be
noted that a similar three peaks distribution was observed in a quite
different task where the primate had to chose between two targets
(both rewarded) presented with a 50ms offset \citep{Schiller2004}.

\subsection{Exploration}

Noise is necessary in the system to allow the generation of saccades
towards one target among two with similar predicted values, rather
than systematically resulting in averaging saccades.  While averaging
saccades sometimes happen in behaving animals \citep{Ottes1984} they
are quite rare and not as systematic as our model would produce them
without perceptual noise. This is because the output of our BG do not
represent a probability distribution of possible targets but indeed a
direct control that requires a unique choice.  Yet, our solution is
probably a bit simplistic, a more plausible one would be to produce a
selection with more competition between targets such as ``race
models'' \citep{Bundesen1987,Ludwig2007}.  These mechanisms would most
of the time allow a selection of a unique target between two perfectly
identical cues.  Moreover these mechanisms could also produce an
attentional engagement/disengagement behavior which could produce the
``gap effect'' \citep{Saslow1967,Braun1988} that our model cannot
replicate.

\subsection{Multiple loops}

In our model we have chosen to include only one SC-Th-BG loop but
\cite{McHaffie2005} have identified at least two (maybe three)
different loops involving different layers of the SC.

The first one linking the SC superficial layers (SCs) to the BG via
lateral posterior (LP) and pulvinar nuclei of thalamus and ending back
to the SC superficial layers (and possibly also deep layers).
According to the fact that SCs activity is mainly driven by direct
retinal projection, it seems reasonable to think that this loop could
be responsible of selection of these retinal inputs.  We didn't
implement this loop that appeared redundant in our model as we
included a SCs to SCi projection but we can imagine a different
mechanism with for example a SCs to SCi pathway gated by SNr
inhibition.

The second loop -- that we implemented in our model -- links the SC
deep layers (SCi) to BG via intralaminar thalamus nuclei (both caudal
and rostral, which represent segregated regions with different type of
contact to striatal medium spiny neurons and thus may in fact describe
two parallel loops).  The deep layers of the SC are known to receive
afferent connections from multiple areas (sensory, premotor, motor,
but also multisensory\dots) \citep{May2006} thus probably conveying
much higher level information.  Moreover, as good evidences
indicate a SCs to SCi projection \citep{Lee1997a,Isa2002}, it seems
reasonable to think that this loop could be involved in selection of
sensory (or high order) targets for orienting behavior as described in
this work.

\subsection{Associative map}

The conjunction task clearly requires the ability to select and
combine feature and location, but we built our model with the
conservative assumption that these different types of information were
treated independently by strictly separating feature and spatial loops
in the learning stage. We thus stick to the assumption of parallel
functionally segregated loops as described in \citep{Alexander1986}.
Moreover this choice was also driven by anatomical considerations as
the TE region of IT seems to projects to the ``Visual Striatum''
\citep{Middleton1996a} while the FEF seems to project to the
``Oculomotor Striatum'' \citep{Stanton1988}. 
This architecture should make learning quicker
and learning generalization easier (i.e. we can directly learn that a
color is rewarded regardless of its position rather than learn each
color/location combination).
This assumption has also the clear advantage to keep the system simple
without the need to learn all possible combinations of features and
locations which would causes a problem of combinatorial explosion.
But the disadvantage is that the system has
no means to directly associate the couple feature/location and can
only separately learn both, explaining the relatively poor
performances for this task.

We hoped that each loop could learn to select
separately and then produce the desired behavior while combined back
at SC level.
However, with this architecture the only mean to perform the correct
behavior (trigger a saccade only if the good cue appears at the good
position) is by triggering an average saccade between the two cues
in the ``no conjunction'' case and thus keeping fixation close to the
center.
In our model, it becomes less and less probable as learning
progresses, because the spatial loop becomes quicker than the feature
one, thus feature information cannot be included in the decision
anymore. Notice that with an external brake (such as inhibition from
dlPFC) limiting the expression of express saccades, the task could
probably be learned.

Different architectures can be proposed to alleviate this problem in
more realistic ways.
It is possible to combine all the information at different levels.
FEF is known to receive inputs from multiple areas
\citep{Schall1995b}, being a convergence structure for ventral and dorsal
visual stream. In particular in our case, IT (TE) is known to project to FEF
\citep{Schall1995b} and we can imagine that FEF already combines
spatial and non-spatial information. This combination could occur
after feature selection and then explain the observed salience map
\citep{Thompson2001}.

Another possibility could be a combination at the Striatum level
allowing the possibility to learn combination of inputs as done in
\citep{Guthrie2013}.  The disadvantage is to multiply the size of the
input vector as stated above. If we have $N$ spatial channels and $M$
color channels the input size is $N \times M$ and the all-to-all
weight matrix $(N \times M)^2$. Even if \cite{Guthrie2013} invoked
interesting biological bases, one can question if this kind of
combination is a problem in biological systems.
The predictions we make about the conjunction case could help deciding
based on experimental data, which architecture (separated or merged
loops) is correct.  

Finally, interaction between loops can also happen at the Thalamus
level.
Even if FEF and IT loops doesn't share the same Thalamic nuclei (VAmc
for IT and MDpl for FEF) this mechanism could still be possible.

\section*{Disclosure/Conflict-of-Interest Statement}

The authors declare that the research was conducted in the absence of
any commercial or financial relationships that could be construed as a
potential conflict of interest.

\section*{Acknowledgement}

\paragraph{Funding:}
This research is funded by the HABOT project (Emergence(s) Ville de Paris program).

\section*{Supplemental Data}

\subsection*{Model parameters}


\begin{center}
\begin{table}[!ht]
\caption{Parameters of the BG model in the spatial loop. The two
  independent Thalamus modules share the same parameters.\label{tab:BG_spatial_params}}
{\begin{tabular}{llllllllll}\hline
     $N$ & $630$ & $\tau$ & $10ms$ & $\tau_{STN}$
     & $5ms$ & $\tau_{FS}$ & $5ms$ & $\tau_{FC}$ & $10ms$
      \\
     \hline
     $\tau_{TH}$ & $5ms$ & $\tau_{TRN}$ & $5ms$ & $\gamma$ &
     $0.2$ & $\W{GPe}{D2}$ & $0.6$ & $\W{D2}{GPe}$ & $0.8$
     \\
     \hline
      $\W{GPe}{D1}$ & $0.6$ & $\W{D1}{GPe}$ & $0.0001$ & $\W{GPe}{FS}$ & $0.001$ &
     $\W{FS}{D1}$ & $0.1$ & $\W{FS}{D2}$ & $0.1$
     \\
     \hline
     $\W{STN}{GPe}$ & $ 0.0003$ &
     $\W{GPe}{STN}$& $0.0003$ & $\W{GPe}{GPi}$ & $0.0002$ & $\W{STN}{GPi}$ & $0.0003$ &
     $\W{D1}{GPi}$ & $0.8$
     \\
     \hline
     $\W{TRN}{TH}$ & $0.003$ & $\W{TH}{TRN}$ & $0.003$ &  $\W{FCtx}{TH}$ & $0.5$ &
     $\W{TH}{FCtx}$ & $3.0$ & $\W{FCtx}{TRN}$ & $0.5$
     \\
     \hline
     $\W{GPi}{TH}$ & $0.9$ & $\W{FCtx}{STN}$ & $1.0$ & $\W{FCtx}{D1}$ &
     $0.1$ & $\W{FCtx}{D2}$ & $0.1$ & $\W{FCtx}{FS}$
     & $0.001$
     \\
     \hline
      $I_{D1}$ & $-0.1$ & $I_{D2}$ & $-0.1$
     & $I_{STN}$ & $0.3$ & $I_{GPi}$ & $0.3$ & $I_{GPe}$
     & $ 0.3$
     \\
     \hline

     $I_{Th}$ & $0.1$& $\W{Input}{D1/D2}$ & $0.9$ & $\W{Input}{FS}$ &
     $0.009$ & $\W{Input}{FC}$ & $0.28$\\ 
     \hline
   \end{tabular}}{}
\end{table}
\end{center}

\begin{center}
\begin{table}[!ht]
\caption{Parameters of the BG model in the color loop. \label{tab:BG_color_params}}
{\begin{tabular}{llllllllll}\hline
     $N$ & $3$ & $\tau$ & $10ms$ & $\tau_{STN}$
     & $5ms$ & $\tau_{FS}$ & $5ms$ & $\tau_{FC}$ & $10ms$
      \\
     \hline
     $\tau_{TH}$ & $5ms$ & $\tau_{TRN}$ & $5ms$ & $\gamma$ &
     $0.2$ & $\W{GPe}{D2}$ & $1.0$ & $\W{D2}{GPe}$ & $0.4$
     \\
     \hline
      $\W{GPe}{D1}$ & $1.0$ & $\W{D1}{GPe}$ & $0.4$ & $\W{GPe}{FS}$ & $0.05$ &
     $\W{FS}{D1}$ & $0.5$ & $\W{FS}{D2}$ & $0.5$
     \\
     \hline
     $\W{STN}{GPe}$ & $ 0.7$ &
     $\W{GPe}{STN}$& $0.45$ & $\W{GPe}{GPi}$ & $0.08$ & $\W{STN}{GPi}$ & $0.7$ &
     $\W{D1}{GPi}$ & $0.4$
     \\
     \hline
     $\W{TRN}{TH}$ & $0.35$ & $\W{TH}{TRN}$ & $0.35$ &  $\W{FCtx}{TH}$ & $0.4$ &
     $\W{TH}{FCtx}$ & $3.0$ & $\W{FCtx}{TRN}$ & $0.35$
     \\
     \hline
     $\W{GPi}{TH}$ & $0.7$ & $\W{FCtx}{STN}$ & $0.58$ & $\W{FCtx}{D1}$ &
     $0.01$ & $\W{FCtx}{D2}$ & $0.01$ & $\W{FCtx}{FS}$
     & $0.01$
     \\
     \hline
      $I_{D1}$ & $-0.1$ & $I_{D2}$ & $-0.1$
     & $I_{STN}$ & $0.5$ & $I_{GPi}$ & $0.1$ & $I_{GPe}$
     & $ 0.1$
     \\
     \hline

     $I_{Th}$ & $0.1$& $\W{Input}{D1/D2}$ & $0.99$ & $\W{Input}{FS}$ & $0.09$ & $\W{Input}{FS}$ & $0.28$\\ 
     \hline
   \end{tabular}}{}
\end{table} 
\end{center}

\begin{center}
\begin{table}[!ht]
\caption{Parameters of the STT model. \label{table:stt_params}}
{\begin{tabular}{llllllllll}\hline
    $\tau$ & $5ms$ & $\tau_{Sat}$ & $100ms$ & $\epsilon_{OPN}$ & $0.1$ & $\epsilon_{trig}$ & $0.4$ & $\epsilon_{{stop}}$ & $0.5$  \\
    \hline
    $\W{SCi}{LLB}$ & $0.15$ &
    $\W{{OPN}}{{Mot}}$ & $10.0$ &  $\W{{OPN}}{{BN}}$ & $40$ &
    $\W{{Mot}}{{Int}}$ & $0.05$ & $\W{{Sat}}{{Mot}}$ & $6.0$\\ 
    \hline
    $\W{{BN}}{{TN}}$ & $0.05$ & $\W{{BN}}{{MN}}$ &
  $1.52$ & $\W{MN}{\theta}$ & $4.07$ \\
\hline
\end{tabular}}{}
\end{table}
\end{center}

\begin{center}
\begin{table}[!ht] 
\caption{Parameters of the SC integration.\label{table:SC_params} }
  {\begin{tabular}{llllllllll}\hline
    $\tau$ & $2ms$ & $\W{SCs}{SCi}$ & $0.5$ &
    $\W{{FEF}}{{SCi}}$ & $0.3$ &  $\W{{V4|IT}}{{SCi}}$ & $0.3$ &
    $\W{{SCi^{in}}}{{SCi}}$ & $0.8$ \\
    \hline
    $\W{BG_{amp}}{SCi}$ & $0.2$ &
    $\W{{BG_{inhib}}}{{SCi}}$ & $3.5$ &  $\W{{SG_{inhib}}}{}$ & $40$ &
    $GPi|SNR_{rest}$ & $0.25$& $SNR_{rest}$ & $0.25$\\
    \hline
\end{tabular}}{}
\end{table}
\end{center}

\begin{center}
\begin{table}[!ht]
\caption{Parameters of the AC modules.  \label{table:AC_params}}
  {\begin{tabular}{llllll}\hline
    $\gamma_{spatial}$ & $0.995$ & $\eta_{spatial}$ & $0.00007$ &
    $\lambda_{spatial}$ & $0.95$\\
    
    \hline

    $\gamma_{color}$ & $0.995$ & $\eta_{color}$ & $0.048$ &
    $\lambda_{color}$ & $0.95$\\
    \hline
\end{tabular}}{}
\end{table}
\end{center}

\bibliographystyle{apalike}
\bibliography{biblio}

\end{document}